\documentclass[final]{elsarticle}
\usepackage[colorlinks]{hyperref}  
\hypersetup{pdfpagemode=None,pdfstartview=FitH}
\usepackage{fullpage}
\usepackage{amsmath}
\usepackage{amsbsy}
\usepackage{amssymb}
\usepackage{amsthm}
\usepackage{amscd}
\usepackage{amsfonts}
\usepackage{supertabular}
\usepackage{yfonts}
\usepackage{graphics}
\usepackage{graphicx}
\usepackage{verbatim}
\usepackage{subfigure}
\usepackage{epsfig}
\usepackage{xspace}
\usepackage{euscript}
\usepackage{alltt}
\usepackage{float}
\usepackage{dsfont}
\usepackage{mathrsfs}
\usepackage{bm}
\usepackage{color}

\newcommand{\tblue}[1]{\textcolor[rgb]{0.0,0.0,0.0}{#1}}

\makeatletter 
\newcommand*\superimpose[2]{%
	\ooalign{$\m@th#1\@firstoftwo#2$\cr
		\hidewidth$\m@th#1\@secondoftwo#2$\hidewidth}%
}
\makeatother
\newcommand*\threedotsord{\mathpalette\superimpose{{\mathop:}{\cdot}}}
\journal{Journal of Computational Physics}
\usepackage{tensor}

\begin{document}

\begin{frontmatter}
	
	
	
	
	\title{Approximation of tensor fields on surfaces of arbitrary topology based on local Monge parametrizations}
	
	
	\author[label1]{Alejandro Torres-S\'anchez}
	\ead{alejandro.torres.sanchez@upc.edu}
	\author[label1]{Daniel Santos-Oliv\'an}
	\author[label1,label2]{Marino Arroyo}
	\ead{marino.arroyo@upc.edu}
	
	\address[label1]{LaC\`aN, Universitat Polit\`ecnica de Catalunya-BarcelonaTech, 08034 Barcelona, Spain}
	\address[label2]{Institute for Bioengineering of Catalonia, The Barcelona Institute of Science and Technology, 08028 Barcelona, Spain}
	
	\begin{abstract}
We introduce a new method, the Local Monge Parametrizations (LMP) method, to approximate tensor fields on general surfaces given by a collection of local parametrizations, e.g.~as in finite element or NURBS surface representations. Our goal is to use this method to solve numerically tensor-valued partial differential equations (PDE) on surfaces. Previous methods use scalar potentials to numerically describe vector fields on surfaces, at the expense of requiring higher-order derivatives of the approximated fields and limited to simply connected surfaces, or represent tangential tensor fields as tensor fields in 3D subjected to constraints, thus increasing the essential number of degrees of freedom. In contrast, the LMP method uses an optimal number of degrees of freedom to represent a tensor, is general with regards to the topology of the surface, and does not increase the order of the PDEs governing the tensor fields. The main idea is to construct maps between the element parametrizations and a local Monge parametrization around each node. We test the LMP method by approximating in a least-squares sense different vector and tensor fields on simply connected and genus-1 surfaces. Furthermore, we apply the LMP method to two physical models on surfaces, involving a tension-driven flow (vector-valued PDE) and nematic ordering (tensor-valued PDE). The LMP method thus solves the long-standing problem of the interpolation of tensors on general surfaces with an optimal number of degrees of freedom.
	\end{abstract}
	
	
\end{frontmatter}

\section{Introduction}

\subsection{Motivation for the approximation of tensor fields on surfaces}

Tensor fields are fundamental objects in continuum physics. Vector fields, the simplest nontrivial instance of a tensor field, appear as a natural representation of the velocity of material particles \cite{Marsden1994-lj} or the polarization in dielectrics \cite{Jackson2007-nf}. Second-order tensor fields are employed to represent fiber distribution or the shape of microscopic inclusions in composite materials \cite{Holzapfel2000-go}, to parametrize nematic ordering in liquid crystals \cite{De_Gennes1995-la,Doi2013-dq}, or to describe internal variables in plasticity \cite{Holzapfel2000-go}. The steady-state or the time-evolution of these tensorial quantities are governed by tensor-valued partial differential equations (PDEs). A classical way to solve vector- or tensor-valued PDEs numerically in two-dimensional or three-dimensional Euclidean space is through the Finite Element Method (FEM), where the Cartesian components of the tensor field are discretized at the nodes of a mesh and approximated with the help of basis functions. However, as we will show in the next section, this straightforward approximation cannot be applied to the components of a tangential vector or tensor field on a surface without increasing the number of essential degrees of freedom, thus also increasing the computational cost involved in the solution. 

Motivated by increasingly quantitative experiments in soft matter and mechanobiology, there is a need to model and simulate the mechanics of thin structures, viewed as surfaces in Euclidean space, including lipid bilayers \cite{Rahimi2012-sa, Arroyo2009-xz,Sahu2017-nz,Sauer2017-oc,Torres-Sanchez2019-kt}, the cell cortex \cite{Turlier2014-wh,Salbreux2017-ln,Torres-Sanchez2019-kt,Mietke2019-fe} or epithelial monolayers \cite{Ishihara2017-pn,Latorre2018-cc}. In most of these works, vector- and tensor-valued PDEs are solved to describe the time-evolution of the surfaces and the different physical quantities characterizing the material. From a theoretical and computational side, there has also been a growing attention to the formulation and resolution of vector- and tensor-valued PDEs on surfaces including Navier-Stokes equations \cite{Nitschke2012-nh,Reuther2015-qv,Nitschke2017-pv,Reuther2018-ky}, vector Laplacians \cite{Hansbo2019-dy}, models for liquid crystal  \cite{Nitschke2018-ru, Praetorius2018-nx, Nestler2018-el,Nitschke2019-hl}, or viscoelasticity and plasticity formulations \cite{Sauer2019-dv}. 
 
Besides finite element simulations of continuum systems, tensor fields on surfaces also play an important role in computer graphics. Vector and tensor fields are designed \cite{Fisher2007-cn,Zhang2006-du,Chen2007-uo,Palacios2007-cf} and routinely discretized \cite{Polthier2003-mk, Tong2003-sr,Hirani2003-ho,Wardetzky2006-gb,Arnold2006-nk,Desbrun2008-ot,Azencot2013-nf,Azencot2015-ho,De_Goes2014-nj} to control the appearance of surfaces, including texture synthesis \cite{Turk2001-ed,Wei2001-va}, line illustration, non-photorealistic rendering \cite{Hertzmann2000-pz}, and anisotropic meshing \cite{Panozzo2014-bf}, among many others \cite{Vaxman2016-tp}.

A few methods exist for the discretization of vector and tensor fields on surfaces in the context of computational mechanics, such as the Hodge decomposition of vector fields for simply-connected surfaces \cite{Nitschke2012-nh,Bhatia2013-ye,Reuther2015-qv,Torres-Sanchez2019-kt,Nitschke2017-pv}, or methods relying on the Cartesian representation of vector and tensor fields mapped to tangential fields using projections \cite{Reuther2018-ky,Fries2018-lv,Nitschke2018-ru,Nestler2019-gk}. In computer graphics, vector fields are discretized as piece-wise constant per face \cite{Polthier2003-mk, Tong2003-sr,Wardetzky2006-gb}, per edge \cite{Hirani2003-ho,Arnold2006-nk,Fisher2007-cn,Desbrun2008-ot}, per vertex \cite{Zhang2006-du,Liu2016-eq}, or encoded as linear operators \cite{Azencot2013-nf,Azencot2015-ho}. However, these methods present several drawbacks for the resolution of vector- or tensor-valued PDEs on surfaces: face-based representations lack of a clear notion of covariant derivative, edge-based discretizations are difficult to generalize to higher-order tensors, and existing vertex-based representations depend on the definition of an alternative notion of covariant derivative (connection) which does not match that of the discretization. We refer to \cite{De_Goes2016-dd} for a thorough and illustrative review of these different kind of discretizations in computer graphics and their shortcomings.
With the goal of solving tensor-valued PDEs on surfaces, we propose a new method based on local Monge parametrizations for the discretization of tensor field on surfaces. We show the application of our methodology in the context of  subdivision surfaces \cite{Stam1999-fk,Cirak2000-tq} and apply it to numerically approximate different problems relevant to soft matter and  biological physics.

\subsection{Description of the problem\label{secDes}}

Our starting point is a surface $\Gamma$ described by piece-wise parametrizations as in finite element approximations, which may be obtained either as an approximation to a given surface or computationally from a physical model. We consider a control mesh consisting on $N_E$ elements, $N_P$ control points and the basis functions $N\indices{_{[E,I]}}(\bm{\xi})$ of control point $I$ for element $E$ taking values in a reference element $\Xi$. The discretization of $\Gamma$ is then given by a collection of finite element parametrizations 
\begin{equation}
\label{finelpar}
\bm{\psi}\indices*{_{[E]}}\left(\bm{\xi}\right) = \sum_{I\in\langle E\rangle}  \bm{x}_{[I]} N\indices{_{[E,I]}}(\bm{\xi}),
\end{equation}
where $\langle E\rangle$ refers to the nodes contributing to the approximation in element $E$ and $\bm{x}\indices{_{[I]}}$ denotes the $I$th control point. The collection of parametrized elements $\Gamma\indices{_{[E]}}=\bm{\psi}\indices{_{[E]}}(\Xi)$ conforms the numerical surface $\Gamma$. We also define $\Gamma_{[I]} = \cup_{E\in\langle I\rangle}\Gamma\indices{_{[E]}}$, where now $\langle I\rangle$ identifies the elements in the neighborhood of a node (the dual of $\langle E\rangle$). 
Let us now consider the discretization of a scalar field $\alpha$ on $\Gamma$, given by the nodal coefficients $\alpha_{[I]}$
\begin{equation}
\label{apprscalar}
\alpha(\bm{x}) = \sum_{I\in\langle E\rangle} \alpha\indices{_{[I]}} N\indices{_{[E,I]}}\circ\bm{\psi}\indices*{_{[E]}^{-1}}(\bm{x}).
\end{equation}
Note that to evaluate $\alpha$ on $\bm{x}\in\Gamma$, we need to compose the basis functions, taking values in the reference element $\Xi$, with the inverse of $\bm{\psi}\indices*{_{[E]}}$. For points $\bm{x}$ on $\Gamma$ belonging to edges between elements, there will be several elements whose image by $\bm{\psi}\indices*{_{[E]}}$ contains $\bm{x}$. Thus, one can compute several different $\bm{\psi}_{[E]}^{-1}$ and calculate Eq.~\eqref{apprscalar} in different elements. However, the resulting $\alpha(\bm{x})$ will be the same in each of these cases provided that the basis functions $N\indices{_{[E,I]}}$ are continuous across elements, i.e.~$N\indices{_{[E,I]}}\circ\bm{\psi}\indices*{_{[E]}^{-1}}(\bm{x}) = N\indices{_{[E',I]}}\circ\bm{\psi}\indices*{_{[E']}^{-1}}(\bm{x})$ whenever $\bm{x}\in \Gamma_{[E]}\cap \Gamma_{[E']}$.

Now, let us consider a vector field $\bm{v}$ of the tangent bundle of $\Gamma$, $T\Gamma=\cup_{\bm{x}\in\Gamma} T_{\bm{x}}\Gamma$ with $T_{\bm{x}}\Gamma$ the tangent plane to $\Gamma$ at $\bm{x}$. Given an orthonormal basis of Euclidean space $\{\bm{i}_1,\bm{i}_2,\bm{i}_3\}$ we could discretize the components of $\bm{v}=v^\alpha \bm{i}_\alpha$ in this basis as
\begin{equation}
v\indices{^{\alpha}}=\sum_{I\in \langle E \rangle}  v\indices{_{[I]}^\alpha}  N\indices{_{[E,I]}} \circ\bm{\psi}\indices*{_{[E]}^{-1}}(\bm{x}), \;\; \alpha = 1,2,3.
\end{equation}
Note that we use Einstein summation convention for repeated indices, denote with greek letters indices running from 1 to 3, and use super-indices for contravariant components of tensors and sub-indices for covariant components.
Because $\bm{v}$ is tangent to $\Gamma$, its three components $v^1,v^2$ and $v^3$ are not independent and need to satisfy that $\bm{v}\cdot\bm{n}=0$, where $\bm{n}$ is the surface normal. In calculations, one would usually need to introduce this additional constraint through a Lagrange multiplier or a penalty \cite{Reuther2018-ky,Fries2018-lv,Nestler2019-gk}.
Alternatively, one could consider the discretization of the components of $\bm{v}=v^a\bm{e}_a$ in a basis $\bm{e}_a$ of the tangent space of $\Gamma$, i.e.
\begin{equation}
\label{tancomp1}
v\indices{^{a}}(\bm{x})=\sum_{I\in \langle E \rangle} v\indices{_{[I]}^a}  N\indices{_{[E,I]}}\circ\bm{\psi}\indices*{_{[E]}^{-1}}(\bm{x}), \;\; a = 1,2.
\end{equation}
Subsequently, we use latin letters to denote indices running from 1 to 2. Eq.~\eqref{tancomp1} requires that the basis $\bm{e}_a$ tangent to $\Gamma$ is defined everywhere. Furthermore, if $\bm{v}$ is to be continuous, then $\bm{e}_a$ also needs to be continuous on $\Gamma$ so that
$\bm{v} = v^a\bm{e}_a$ is continuous. 
However, one cannot define such a basis for a general closed surface, e.g.~on a sphere, as a consequence of the hairy ball theorem \cite{Milnor1978-gf}. For instance, polar coordinates in the sphere present singularities at the poles. \tblue{We also note that the natural basis of the tangent space given by the finite element parametrizations and defined  by the vectors $\partial\indices{_1}\bm{\psi}\indices*{_{[E]}}$ and $\partial\indices{_2}\bm{\psi}\indices*{_{[E]}}$ is discontinuous across elements due to the jump in the definition of local coordinates from one element to another. Thus, while these bases would be the natural choice in a FEM setup, an approximation of the type  $\bm{v}=\sum_{I\in \langle E \rangle}v\indices{_{[I]}^a}\partial\indices{_a}\bm{\psi}\indices*{_{[E]}}$, where $v\indices{_{[I]}^a}$ are nodal coefficients for the components of $\bm{v}$,  is discontinuous across elements due to the discontinuity of $ \partial\indices{_1}\bm{\psi}\indices*{_{[E]}}$ and $\partial\indices{_2}\bm{\psi}\indices*{_{[E]}}$. These are fundamental difficulties to an approximation of the tangential components of a vector field as in Eq.~\eqref{tancomp1}.}

\subsection{Hodge decomposition}

A classical approach to discretize a vector field on a surface using only tangential calculus rather than  relying on its interdependent three-dimensional components is the so-called Hodge decomposition of a vector field in terms of scalar potentials, which are trivial to approximate following Eq.~\eqref{apprscalar}. According to this decomposition, a vector field $\bm{v}$ tangent to a surface $\Gamma$ can be represented as
\begin{equation}
\label{velhodge}
\bm{v} = \nabla \alpha + \nabla\times \beta + \bm{h},
\end{equation}
where $\alpha$ and $\beta$ are scalar fields on $\Gamma$ and $\bm{h}$ is a harmonic vector field, satisfying $\nabla\cdot\bm{h}=0$ and $\nabla\times\bm{h}=0$.  Here, $\nabla$ denotes the surface gradient or covariant derivative. Note that the curl operator $\nabla\times$ on a surface (an instance of exterior derivative) applied on a scalar function $\beta$ is a vector with components $(\nabla\times \beta)^a = \epsilon^{ab} \nabla_b\beta$, where $\bm{\epsilon}$ is the antisymmetric tensor, with components
\begin{equation}
\epsilon^{ab} = J^{-1} \varepsilon^{ab}, 
\end{equation}
with $\varepsilon^{ab}$ the Levi-Civita symbols, defined by $\varepsilon^{11} = \varepsilon^{22} = 0$, $\varepsilon^{12} = -\varepsilon^{21} = 1$, $J=\sqrt{\det\bm{g}}$ and $\bm{g}$ is the metric tensor on the surface, whose components in a basis $\{\bm{e}_1,\bm{e}_2\}$ of $T\Gamma$ are given by $g_{ab} = \bm{e}_a\cdot\bm{e}_b$.  When acting on a vector, $\nabla\times$ leads to a scalar $\nabla\times\bm{h}= \epsilon\indices{^b_a} \nabla\indices{_b} h\indices{^a}$. It is not clear how Eq.~\eqref{velhodge} could help us discretize a vector field since,  while the scalar fields $\alpha$ and $\beta$ are easily dealt with, one still needs to approximate the harmonic vector field $\bm{h}$. For simply connected surfaces, i.e.~closed surfaces with genus equal to 0 topologically equivalent to a sphere, there is only a trivial harmonic vector field, $\bm{h}=\bm{0}$, and thus $\bm{v}$ can be described in terms of the two scalar fields $\alpha$ and $\beta$. In this case, one can approximate $\alpha$ and $\beta$
\begin{equation}
\alpha(\bm{x}) = \sum_{I\in\langle E\rangle} \alpha\indices{_{[I]}} N\indices{_{[E,I]}}\circ\bm{\psi}_{[E]}^{-1}(\bm{x}),\qquad \beta(\bm{x}) = \sum_{I\in\langle E\rangle} \beta\indices{_{[I]}} N\indices{_{[E,I]}}\circ\bm{\psi}\indices*{_{[E]}^{-1}}(\bm{x}),
\end{equation}
and consider the discretization of $\bm{v}$ as
\begin{equation}
\bm{v}(\bm{x}) = \sum_{I\in\langle E\rangle}\left[\left( \alpha_{[I]} g\indices{^{ab}}\nabla\indices{_b} N\indices{_{[E,I]}} + \beta_{[I]} \epsilon\indices{^{ab}}  \nabla\indices{_b} N\indices{_{[E,I]}}\right)\partial\indices{_a} \bm{\psi}\indices*{_{[E]}} \right]\circ\bm{\psi}\indices*{_{[E]}^{-1}}(\bm{x}).
\end{equation}
This method, which we refer to as the Hodge Decomposition (HD) method, was introduced in the context of fluid interfaces by Secomb and Skalak \cite{Secomb1982-cf} and subsequently used by many others, mostly for inextensible flows on stationary surfaces where $\bm{v}=\nabla\times\beta$ (since $\nabla\cdot\bm{v} = \Delta\alpha = 0$) \cite{Nitschke2012-nh,Morris2015-up,Reuther2015-qv,Sigurdsson2016-jt,Gross2018-nt,Mickelin2018-ug}. 
More recently, its full representation involving both $\alpha$ and $\beta$ has been used to solve flows on surfaces with imposed evolution \cite{Reuther2015-qv} or in self-consistent models including equations for  shape evolution \cite{Torres-Sanchez2019-kt}. 
In computer graphics, the HD is routinely used for vector field design \cite{Bhatia2013-ye,De_Goes2016-dd}.
However, this method is restricted to simply connected surfaces and cannot be applied to higher-order tensor fields, although an extension for second-order tensor fields based on discrete exterior calculus has been recently proposed \cite{De_Goes2014-nj}. Furthermore, as the representation of the vector field already involves derivatives, it leads to higher-order PDEs. In the next section, we introduce a new method for the approximation of general tensor fields on surfaces of arbitrary topology.

\section{Description of the LMP method}

Discretizing tensor fields for the resolution of tensor-valued PDEs on surfaces of arbitrary topology based only on tangential calculus remains an open problem. The main goal of this work is to provide a general method to address this issue, which we call the Local Monge Parametrizations (LMP) method. The  LMP method can be synthesized in three steps. First, one needs to construct a set of local Monge parametrizations of $\Gamma$ around each node of the mesh. Then, one needs to find the changes of coordinates between local Monge parametrizations and the finite element bases. We describe how to find a suitable set of local Monge parametrizations and how to compute the changes of basis in section \ref{locmonpar}. In the LMP method, the components of a tensor are discretized at the nodes of the mesh in terms of their respective local Monge parametrizations. As we describe in section \ref{intermeth}, the LMP method then uses the changes of basis from local Monge parametrizations to the finite element parametrizations to reconstruct the components of the tensor in the finite element parametrization used for calculations. 
 
\subsection{Local Monge parametrizations}
\label{locmonpar}

\begin{figure}
\begin{center}
\includegraphics[width=6in]{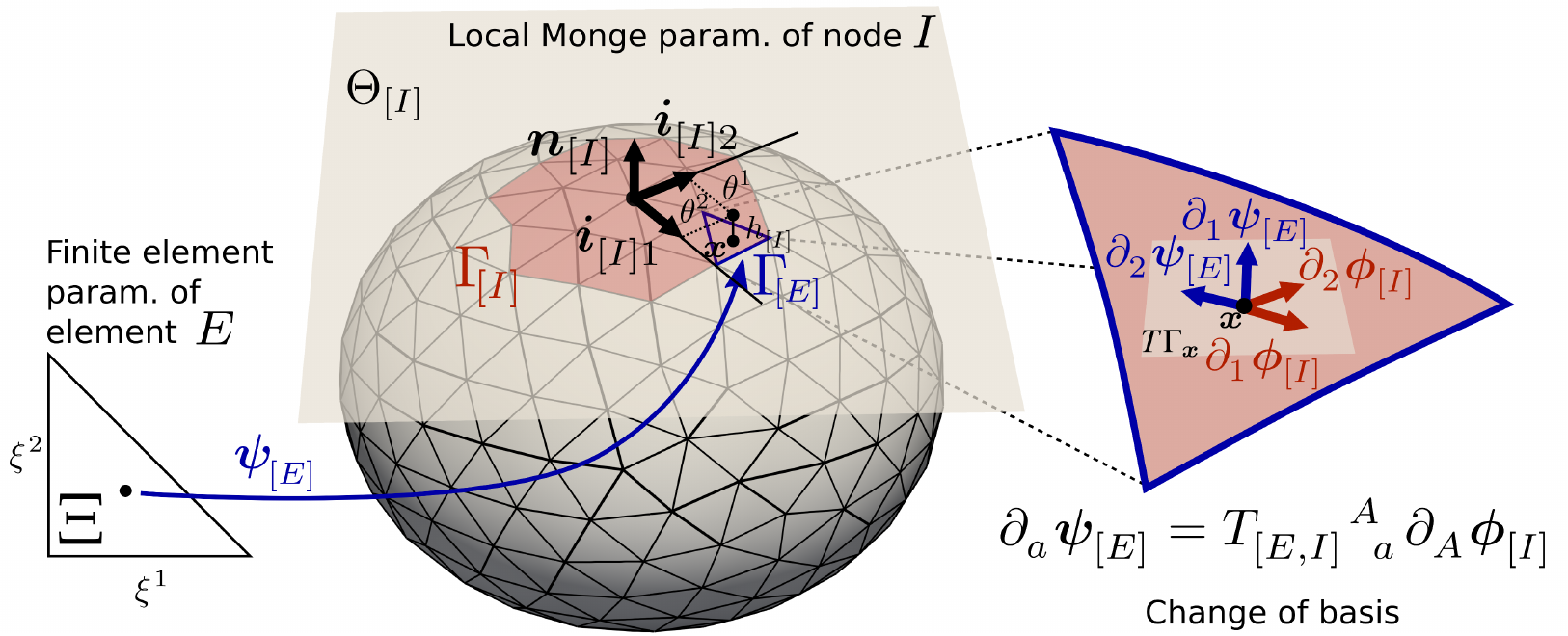}
\caption{\label{fig_method} Local Monge parametrization. In the neighborhood of node $I$, defined as the support of the basis function associated with it (light red), a Monge parametrization relative to a plane $\Theta\indices{_{[I]}}$ passing through $\bm{x}\indices{_{[I]}}$ can be defined. The two coordinates $\theta^1$ and $\theta^2$ define the position of a point in $\Theta\indices{_{[I]}}$ and parametrize the surface with the help of a height function $h\indices{_{[I]}}$. Given a triangle of the finite element discretization $\Gamma\indices{_{[E]}}$ and its parametrization from the reference element $\Xi$, $\bm{\psi}\indices{_{[E]}}$, one can construct in each point of this element a linear map between the bases of the tangent plane at this point given by $\mathcal{B}_{\bm{\phi}_{[I]}} = \{ \partial_A \bm{\phi}_I\}$ to $\mathcal{B}_{\bm{\psi}_{[E]}} = \{ \partial_a \bm{\psi}_{[E]}\}$.}
\end{center}
\end{figure}

We introduce next an atlas of local Monge parametrizations, one per node of the control mesh. Given  node $I$, we recall that its  neighborhood $\Gamma\indices{_{[I]}}$ is the image of all element parametrizations of the form in Eq.~\eqref{finelpar} to which node $I$ contributes. We further define a plane $\Theta\indices{_{[I]}}$ passing through $\bm{x}\indices{_{[I]}}$, which does not need to be tangent to $\Gamma$ but must satisfy that in $\Gamma\indices{_{[I]}}$ the projection of $\Gamma$ onto $\Theta\indices{_{[I]}}$ is one-to-one. We define a basis of $\Theta\indices{_{[I]}}$ given by the orthonormal vectors $\bm{i}\indices{_{[I]}_1}$ and $\bm{i}\indices{_{[I]}_2}$ and we denote by $\bm{n}\indices{_{[I]}}$ the unit normal to $\Theta\indices{_{[I]}}$ (see Fig.~\ref{fig_method}). Thus, $\{\bm{i}\indices{_{[I]}_1},\bm{i}\indices{_{[I]}_2},\bm{n}\indices{_{[I]}}\}$ forms an orthonormal basis of Euclidean space.

We can now define a local Monge parametrization of the surface around $I$, $\bm{\phi}\indices{_{[I]}}: \Theta\indices{_{[I]}}\rightarrow \Gamma\indices{_{[I]}}$ given by 
\begin{equation}
\label{eqlocmonpar}
\bm{\phi}\indices{_{[I]}}(\theta^{1},\theta^{2}) = \bm{x}\indices{_{[I]}} + \theta^{1} \bm{i}\indices{_{[I]}_1}+ \theta^{2}\bm{i}\indices{_{[I]}_2} + h\indices{_{[I]}}(\theta^{1},\theta^{2}) \bm{n}\indices{_{[I]}},
\end{equation}
where $h\indices{_{[I]}}$ is the height function from $\Theta\indices{_{[I]}}$ to $\Gamma\indices{_{[I]}}$. The basis vectors of the $T_{\bm{x}}\Gamma$ induced by this parametrization are $\partial\indices{_A} \bm{\phi}_{[I]} =  \bm{i}\indices{_{[I]}_A} + \partial\indices{_A} h\indices{_I} \bm{n}\indices{_{[I]}}, \; A=1,2$, where $\partial\indices{_A}$ denotes partial differentiation with respect to $\theta\indices{_A}$, and we use capital indices to refer to the coordinates in $\Theta\indices{_{[I]}}$. At each point in $\Gamma\indices{_{[I]}}$, we have thus two bases of the tangent plane to the surface, that induced by the parametrization of the corresponding element $\mathcal{B}_{\bm{\psi}_{[E]}} = \left\{ \partial_a \bm{\psi}_{[E]}\right\}$, and that of the local Monge parametrization of node $I$ $\mathcal{B}_{\bm{\phi}_{[I]}} = \left\{ \partial_A \bm{\phi}_I\right\}$. We can then express a tangent vector $\bm{v}$ locally in either of these bases 
\begin{equation}
\bm{v} = v\indices{^a} \partial\indices{_a} \bm{\psi}\indices*{_{[E]}} = V\indices{^A} \partial\indices{_A} \bm{\phi}\indices{_{[I]}}.
\end{equation}
The change of basis between $\mathcal{B}_{\bm{\psi}_{[E]}}$ and $\mathcal{B}_{\bm{\phi}_{[I]}}$ is just a linear map $\bm{T}_{[E,I]}$ from the tangent plane to itself given by four coefficients. A simple calculation shows (see \ref{change-basis}) that  
\begin{equation}
\label{transform1}
T\indices{_{[E,I]}^A_a}(\theta_1,\theta_2)  = \bm{i}\indices{_{[I]}_A}\cdot\partial\indices{_a}\bm{\psi}\indices*{_{[E]}}(\theta_1,\theta_2),
\end{equation}
from which we can relate the tangent vectors in $\mathcal{B}_{\bm{\psi}_{[E]}}$ and $\mathcal{B}_{\bm{\phi}_{[I]}}$ as
\begin{equation}
\partial\indices{_a} \bm{\psi}\indices*{_{[E]}} = T\indices{_{[E,I]}^A_a}  \partial\indices{_A} \bm{\phi}\indices{_{[I]}} \;\; \mbox{and}\;\; \partial\indices{_A} \bm{\phi}\indices{_{[I]}} = \hat{T}\indices{_{[E,I]}^a_A} \partial\indices{_a} \bm{\psi}\indices*{_{[E]}},
\label{eq10}
\end{equation}
where $\hat{\bm{T}}\indices{_{[E,I]}} = \bm{T}\indices{_{[E,I]}^{-1}}$. Remarkably, $\bm{T}\indices{_{[E,I]}}$ is independent of $h\indices{_{[I]}}$.

\subsection{Approximation of tensor fields\label{intermeth}}

Let us first describe the interpolation of a vector field $\bm{v}$ by the LMP method and leave the interpolation of higher-order tensors for later in this section. We will approximate a vector field over the surface in element $E$ as a linear combination of the form
\begin{equation}
\bm{v}(\bm{x}) = \sum_{I\in\langle E\rangle} \left(\bm{v}\indices{_{[I]}} \; N\indices{_{[E,I]}}\right)\circ\bm{\psi}^{-1}_{[E]}(\bm{x}).
\label{eq11}
\end{equation}
In a usual FEM discretization, $\bm{v}\indices{_{[I]}}$ would be a constant vector. For general surfaces with Gaussian curvature\tblue{, i.e.~with non-vanishing Riemann tensor,} however, there is a fundamental difficulty to define a constant vector field \cite{Do_Carmo1992-bx,De_Goes2016-dd}. \tblue{For instance, parallel transporting a vector on a surface with Gaussian curvature around a closed loop leads to a twist of the vector}. For this reason, around each node $I$ we define a vector field $\bm{v}\indices{_{[I]}}(\bm{x})$ with support in $\Gamma\indices{_{[I]}}$ and given by 
\begin{equation}
\bm{v}_{[I]}(\bm{x}) = V\indices{_{[I]}^A} \left(\partial_A \bm{\phi}_{[I]} \circ\bm{\phi}^{-1}_{[I]}(\bm{x})\right),
\label{eq12}
\end{equation}
which plays the role of the coefficient of node $I$. In fact, this local vector field is such that its components are constant in the basis $\mathcal{B}_{\bm{\phi}_{[I]}}$ of the $T_{\bm{x}}\Gamma$. This local vector field is continuous provided that $\bm{\phi}\indices{_{[I]}}$, the local Monge parametrization of node $I$, is continuously differentiable, which is the case if $\Gamma\indices{_{[I]}}$ is continuously differentiable. Since all the local Monge parametrizations form an atlas of the surface, this requires $\Gamma$ to be continuously differentiable. If this is the case, since the basis functions $N\indices{_{[E,I]}}$ are continuous across elements, then the element-wise approximations in Eq.~\eqref{eq11} define a continuous vector field over $\Gamma$. Following this argument, a surface defined by $C^0$ finite elements would not lead to a continuous vector field using this method. \tblue{Note, however, that this limitation is not intrinsic to the method  but rather a consequence of the fact that a tangent vector field on a $C^0$ surface is necessarily discontinuous.} Instead, Loop subdivision surfaces are $C^2$ everywhere except at irregular points \cite{Stam1999-fk,Cirak2000-tq}, where they present jumps in second-order derivatives, and hence $\bm{v}$ will be continuously differentiable everywhere except at irregular points.

Plugging Eq.~\eqref{eq12} into \eqref{eq11} and using \eqref{eq10}, we are able to express $\bm{v}(\bm{x})$ fully in terms of quantities computable at the finite-element level as
\begin{equation}
\label{lmp_vector}
\bm{v}(\bm{x}) = \sum_{I\in\langle E\rangle} V\indices{_{[I]}^A} \left(\hat{T}\indices{_{[E,I]}^a_A}  \; N\indices{_{[E,I]}} \; \partial_a \bm{\psi}_{[E]}\right)\circ\bm{\psi}^{-1}_{[E]}(\bm{x}),
\end{equation}
or in components as
\begin{equation}
\label{lmp_vector}
{v}^a(\bm{x}) = \sum_{I\in\langle E\rangle} V\indices{_{[I]}^A} \left(\hat{T}\indices{_{[E,I]}^a_A}  \; N\indices{_{[E,I]}} \right)\circ\bm{\psi}^{-1}_{[E]}(\bm{x}).
\end{equation}
In each element, this vector field is parametrized by the nodal coefficients $V\indices{_{[I]}^A}$ and expressed in the natural basis $\mathcal{B}_{\bm{\psi}_E}$.  Although the basis vectors $\partial\indices{_a} \bm{\psi}\indices*{_{[E]}}$ are discontinuous, the also discontinuous linear transformations $\hat{\bm{T}}\indices{_{[E,I]}}$ make the resulting approximation continuous as discussed above.

This method can be generalized to any kind of tensor field over $\Gamma$, noting that the dual bases of $\mathcal{B}_{\bm{\psi}_{[E]}}$ and $\mathcal{B}_{\bm{\phi}_{[I]}}$ transform analogously to Eq.~\eqref{eq10} but with the inverse matrix. For a 1-form $\bm{\alpha}$, one can thus write its components in the dual of $\mathcal{B}_{\bm{\psi}_{[E]}}$ as
\begin{equation}
\alpha\indices{_a}(\bm{x}) = 
\sum_{I\in\langle E\rangle} \alpha\indices{_{[I]}_A} \left({T}\indices{_{[E,I]}^A_a}  \; N\indices{_{[E,I]}}\right)\circ\bm{\psi}\indices*{_{[E]}^{-1}}(\bm{x}).
\end{equation}
For a general tensor, this can be trivially generalized to 
\begin{equation}
\label{general_tensor}
q\indices{_{a_1}_\dots_{a_m}^{b_1}^\dots^{b_n}}(\bm{x}) = \sum_{I\langle E\rangle}q\indices{_{[I]}_{C_1}_\dots_{C_m}^{D_1}^\dots^{D_n}} \left(T\indices{_{[E,I]}^{C_1}_{a_1}}\cdots T\indices{_{[E,I]}^{C_n}_{a_n}} \hat{T}\indices{_{[E,I]}^{b_1}_{D_1}}\cdots \hat{T}\indices{_{[E,I]}^{b_m}_{D_m}} N\indices{_{[E,I]}}  \right)\circ\bm{\psi}\indices*{_{[E]}^{-1}}(\bm{x}).
\end{equation}

\section{Numerical approximation of vector and tensor fields with LMP and subdivision finite elements}

Here, we consider a FEM setup based on subdivision surfaces \cite{Stam1999-fk,Biermann2000-bz} similar to that in \cite{Torres-Sanchez2019-kt} to examine numerically the approximation power of the LMP method by computing the $L_2$ projection $\bm{v}(\bm{x})$ as in Eq.~\eqref{lmp_vector} of a given vector field $\bm{w}(\bm{x})$ on a given surface.
\begin{figure}
	\begin{center}
		\includegraphics[width=6.5in]{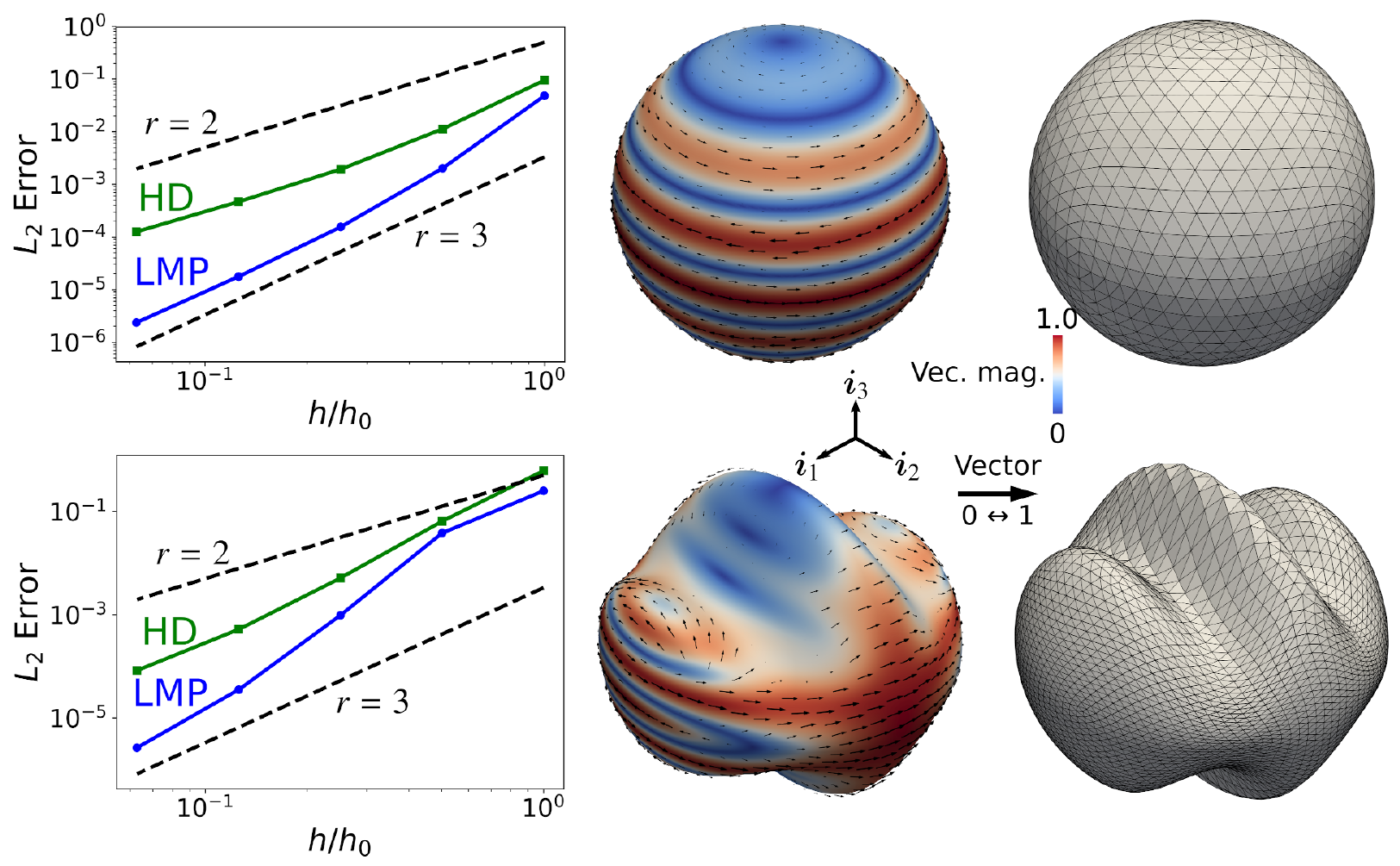}
		\caption{\label{figsph} Comparison between the approximation power of the LMP and HD methods on simply connected surfaces with a discretization based on subdivision surfaces. We observe a convergence rate of the $L_2$ error as a function of element size of $3$ for the LMP method and of $2$ for the HD method for two different surface geometries. In the right, we show the solution found with the LMP method for the finest mesh and the coarser mesh used for geometry.}
	\end{center}
\end{figure}
For a vector field, finding the least-squares fit amounts to solving a linear system
\begin{equation}
\mathsf{K} \mathsf{v} = \mathsf{b},
\end{equation}
where
\begin{equation}
\mathsf{K}_{2I+A,2J+B} = \sum_{E\in\langle I\rangle \cap \langle J\rangle} \int_{\Gamma\indices{_{[E]}}}  g\indices{_{ab}}(\bm{x}) \left(N\indices{_{[E,I]}} N\indices{_{[E,J]}}  \hat{T}\indices{_{[E,I]}^a_A} \hat{T}\indices{_{[E,J]}^b_B} \right) \circ\bm{\psi}\indices*{_{[E]}^{-1}}(\bm{x})\,dS,
\end{equation}
\begin{equation}
\mathsf{b}_{2I+A} = \sum_{E\in\langle I\rangle \cap \langle J\rangle} \int_{\Gamma\indices{_{[E]}}}  w\indices{^b}(\bm{x}) g\indices{_{ab}}(\bm{x}) \left(N\indices{_{[E,I]}} \hat{T}\indices{_{[E,I]}^a_A}  \right) \circ\bm{\psi}\indices*{_{[E]}^{-1}}(\bm{x})\,dS.
\end{equation}
Here, $g\indices{_{ab}}$ are the components of the metric tensor in the basis $\mathcal{B}_{\bm{\psi}_{[E]}}$. To define a vector field $\bm{w}(\bm{x})$ tangent to $\Gamma$, we compute the projection onto $\Gamma$ of a three-dimensional vector field $\bm{W}(\bm{x})$. 
We start by examining the approximation of the vector field $\bm{W}(\bm{x})=\cos(6\pi z)(-y,x,0)$ on a sphere of radius 1 as we refine the mesh size. 
\tblue{We project $\bm{W}$ onto the discrete surface $\Gamma$ and not onto the exact sphere because, by construction, the LMP method leads to a vector field that is tangent to the discrete surface $\Gamma$. Both the projection of $\bm{W}$ onto $\Gamma$ and the solution $\bm{v}$ will converge to the projection of $\bm{W}$ on the exact sphere as $\Gamma$ is refined.} We compute $\Gamma$ by a least-squares fit to the sphere. We find that the field $\bm{v}$ represented using the LMP method converges to $\bm{w}$ with an error scaling as $h^{-r}$, with $h$ being the average mesh size and with rate $r=3$, \tblue{consistent with the approximation power of subdivision surfaces \cite{Arden2001-id}}, see Fig.~\ref{figsph} top. A similar analysis with the HD method, shows that the latter converges with a rate $r=2$. The rate of convergence is unaffected by geometric non-axisymmetric perturbations of the sphere, see Fig~\ref{figsph} bottom. \tblue{To obtain this shape, we perform again a least-squares fit to the deformed sphere parametrized by
$\bm{x}(\varphi,\theta) = \left(
\sin\theta\cos\varphi,
\sin\theta\sin\varphi,
\lambda_1 \cos\theta  (1+\lambda_2\cos(2\pi\sin\theta\sin\varphi))
\right)$, with $\lambda_1 = 0.7$ and $\lambda_2=0.3$.} The faster convergence rate of the LMP method can be expected since HD involves one additional derivative of the basis functions.

\begin{figure}
	\begin{center}
		\includegraphics[width=6.5in]{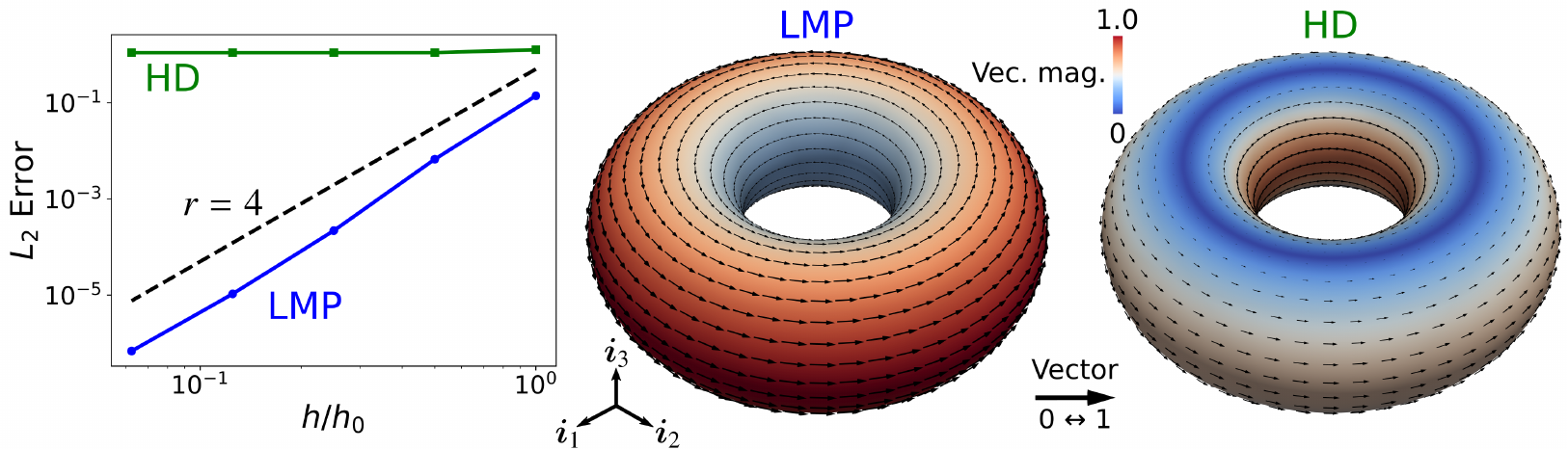}
		\caption{\label{figtor}Comparison between the LMP method and the HD method for the approximation of a vector field $\bm{v}=(-y,0,0)$ on a torus. While the LMP method is able to represent $\bm{v}$ on the torus, the HD method fails because the harmonic part of $\bm{v}$ cannot be represented using vector potentials.}
	\end{center}
\end{figure}

Considering now a non-simply connected surface with genus 1, a torus, Fig.~\ref{figtor} shows that while the LMP method is able to approximate a vector field $\bm{v}$ with non-zero harmonic component, the  HD method fails due to its inability to represent the harmonic component, see \cite{Nitschke2017-pv} for a detailed discussion on this issue. \tblue{We also note that, for the torus in Fig.~\ref{figtor} the rate of convergence of the LMP method increases to four ($r=4$). This is due to the fact that the torus is triangulated without irregular points, i.e.~all vertices in the mesh have six neighbors; the approximation power of subdivision surfaces in the absence of irregular points is four \cite{Arden2001-id}.}

\begin{figure}
	\begin{center}
		\includegraphics[width=6.5in]{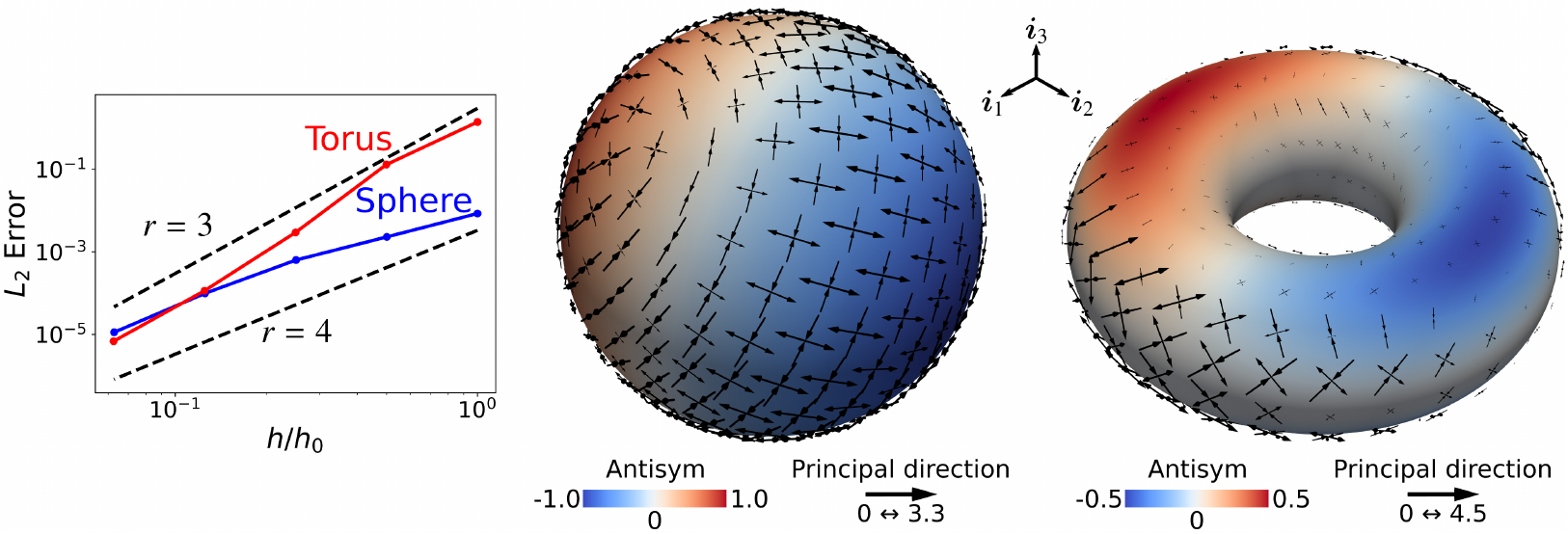}
		\caption{\label{figtens}LMP method applied to the interpolation of general second order tensor fields on a surface. (left) $L_2$ error convergence as a function of mesh size. We observe a cubic convergence for LMP method on a sphere (with irregular points) and a quartic convergence on a torus (no irregular points). (center) Example of tensor field on a sphere reproduced with the LMP method. The antisymmetric component $\epsilon_{ab} \sigma^{ab}$ is plotted as a colormap, while the symmetric part $(\sigma^{ab}+\sigma^{ba})/2$ is represented with two pairs of arrows whose direction indicate the two principal directions of the tensor and whose size is proportional to the (signed) eigenvalues (inward and outward arrows indicate negative and positive eigenvalues respectively). (right) Same plot on a torus.}
	\end{center}
\end{figure}
Finally, we consider the discretization of a general second-order tensor field $\bm{\sigma}$. In this case, the matrix and right-hand sides of the system have the form
\begin{equation}
\mathsf{K}_{4I+2A+B,4J+2C+D}^h = \sum_{E\in\langle I\rangle \cap \langle J\rangle} \int_{\Gamma\indices{_{[E]}}}  g\indices{_{ac}}(\bm{x})g\indices{_{bd}}(\bm{x}) \left(N\indices{_{[E,I]}} N\indices{_{[E,J]}}    \hat{T}\indices{_{[E,I]}^a_A} \hat{T}\indices{_{[E,I]}^b_B} \hat{T}\indices{_{[E,J]}^c_C} \hat{T}\indices{_{[E,J]}^d_D} \right) \circ\bm{\psi}\indices*{_{[E]}^{-1}}(\bm{x}) dS,
\end{equation}
\begin{equation}
\mathsf{b}_{4I+2A+B}^h = \sum_{E\in\langle I\rangle \cap \langle J\rangle} \int_{\Gamma\indices{_{[E]}}}  \sigma\indices{^{cd}}(\bm{x}) g\indices{_{ca}}(\bm{x})g\indices{_{db}}(\bm{x})\left(N\indices{_{[E,I]}} \hat{T}\indices{_{[E,I]}^a_A} \hat{T}\indices{_{[E,I]}^b_B}  \right) \circ\bm{\psi}\indices*{_{[E]}^{-1}}(\bm{x}) dS.
\end{equation}
To test the convergence of the method, we consider the three dimensional tensor $\bm{\Sigma} = ((0, -1, -1), (-1 , 0, 1),\\ (-1, 0, -1))$, and compute its projection onto $\Gamma$, $\sigma\indices{^{ab}} = (\bm{e}\indices{^a}\cdot\bm{i}\indices{_\alpha}) (\bm{e}\indices{^b}\cdot\bm{i}\indices{_\beta}) \Sigma\indices{^{\alpha\beta}}$. Again, we find an approximation power of $r=3$ for a sphere and $r=4$ for a torus (no irregular points), see Fig.~\ref{figtens}. To plot the tensor field, we first decompose its antisymmetric part $\epsilon\indices{_{ab}}\sigma\indices{^{ab}}$, a scalar on $\Gamma$, and its symmetric part $(\sigma\indices{^{ab}}+\sigma\indices{^{ba}})/2$. We plot the antisymmetric part as a colormap. For the symmetric part, we compute its eigenvalues and eigenvectors and plot two pairs of arrows at every point denoting the direction of the eigenvectors with a magnitude equal to the (signed) magnitude of the eigenvalues (inward arrows indicate negative eigenvalues whereas outward arrows indicate positive ones).
\section{Numerical solution of vector- and tensor-valued PDEs on surfaces with the LMP method}
\subsection{Tension-driven flow}
In this section, we apply our methodology to a tension-driven flow on a surface, described by the PDE
\begin{equation}
\label{marangoni}
2\mu\nabla\cdot\nabla^S\bm{v} + \nabla\gamma = \eta \bm{v}.
\end{equation}
The first term in the left-hand side of the previous equation represents the divergence of a viscous stress $2\mu\nabla^S\bm{v}$, where $(\nabla^S \bm{v})_{ab}=(\nabla_av_b+\nabla_bv_a)/2$ is the symmetric gradient of $\bm{v}$ and $\mu$ the shear viscosity, whereas the second term represents the gradient of a surface tension $\gamma$. The right-hand side represents friction with the surrounding medium characterized by the friction coefficient $\eta$. 
\begin{figure}
	\begin{center}
		\includegraphics[width=5in]{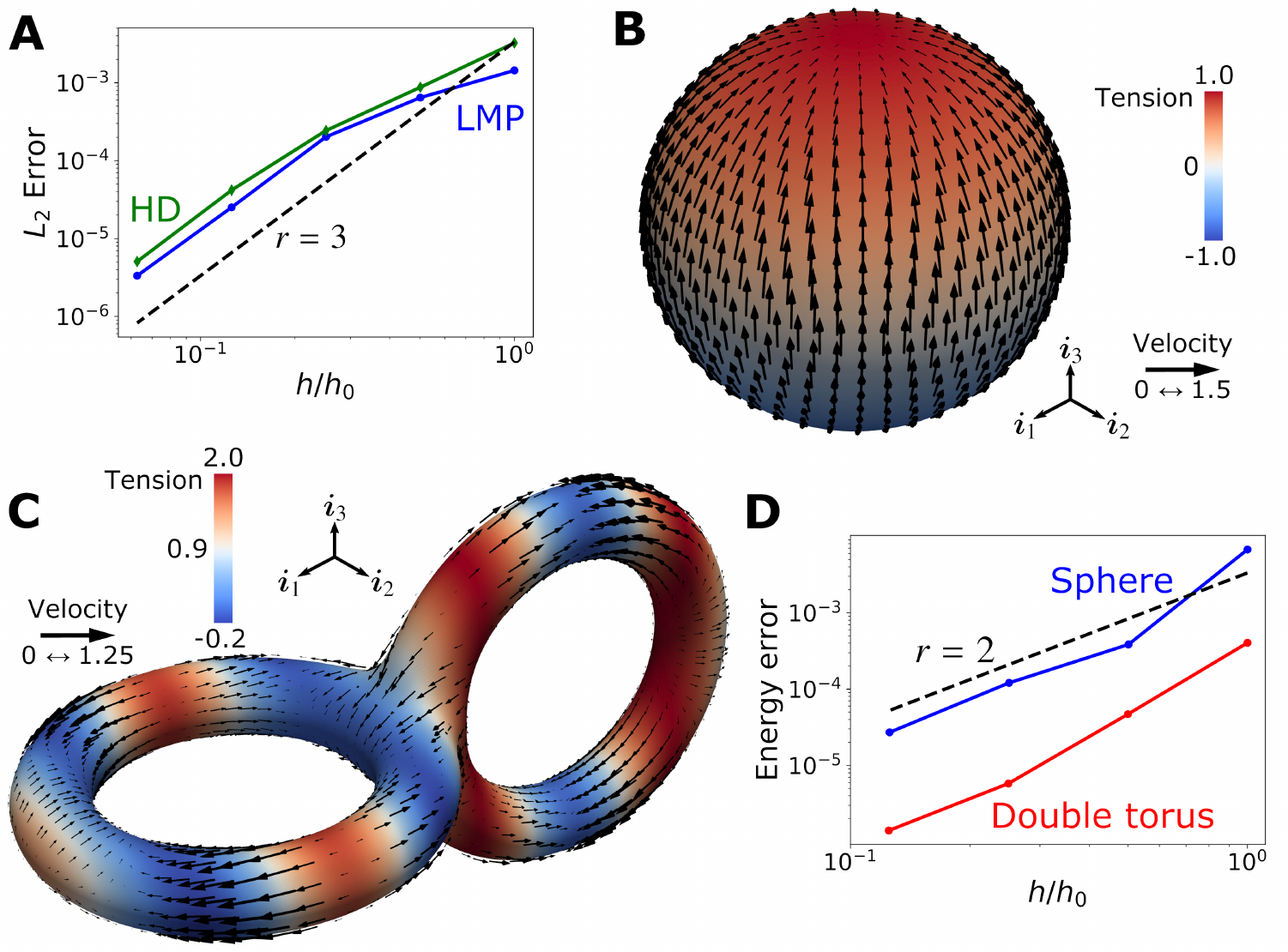}
		\caption{\label{figmarfl} \tblue{(A and B) Convergence of a tension-driven flow on a sphere with a tension gradient $\gamma=z$. We observe, again, a cubic convergence for the LMP method and a better performance of the HD method that for previous cases, probably caused by a superconvergence due to the lack of solenoidal field in this example. (C) LMP method applied on a double torus (see main text) with $\gamma=\cos(3\pi x) + \exp(-y^2)$. (D) In the energy norm, we observe that the LMP method convergences with a quadratic convergence irrespective of the topology.}}
	\end{center}
\end{figure}
A similar equation, but allowing for shape deformations, can be shown to describe actin flows on the cell cortex in the limit of very fast turnover \cite{Torres-Sanchez2019-kt,Turlier2014-wh,Mietke2019-fe}. Here, we restrict ourselves to non-evolving surfaces for simplicity, but combining the LMP method with the procedure described in \cite{Torres-Sanchez2019-kt}, where we used the HD method instead, is \tblue{simple, as we show in the discussion section}. The unknown in this problem is the field $\bm{v}$. Here, we discretize $\bm{v}$ in terms of its covariant components using the LMP representation
\begin{equation}
v\indices{_a}(\bm{x}) = \sum_{I\in\langle E\rangle} v\indices{^I_A} N\indices{_{[E,I]}} T\indices{_{[E,I]}^A_a} \circ\bm{\psi}\indices*{_{[E]}^{-1}}(\bm{x}).
\end{equation}
To solve this problem numerically, we consider a Garlerkin method and multiply Eq.~\eqref{marangoni} by the ``tensorial'' test functions
\begin{equation}
u\indices{^I_a}(\bm{x}) = N\indices{_{[E,I]}} T\indices{_{[E,I]}^1_a} \circ\bm{\psi}\indices*{_{[E]}^{-1}}(\bm{x}), \qquad w\indices{^I_a}(\bm{x}) =N\indices{_{[E,I]}} T\indices{_{[E,I]}^2_a} \circ\bm{\psi}\indices*{_{[E]}^{-1}}(\bm{x}).
\end{equation}
After integration by parts, we obtain the linear system of equations
\begin{equation}
\mathsf{K}\mathsf{v} = \mathsf{b}, 
\end{equation}
where
\begin{equation}
\mathsf{K}_{2I+a,2J+b}= \sum_E \int_{\Gamma\indices{_{[E]}}}\left[2\mu  \nabla\indices{_b}(N\indices{_{[E,I]}} T\indices{_{[E,I]}^A_a}) g\indices{^{ac}}g\indices{^{bd}} \nabla\indices{_d}(N\indices{_{[E,J]}}T\indices{_{[E,J]}^B_b}) + \frac{\eta}{2} N\indices{_{[E,I]}} N\indices{_{[E,J]}} T\indices{_{[E,I]}^A_a}g^{ab}T\indices{_{[E,J]}^B_b}\right]dS,
\end{equation}
and
\begin{equation}
\mathsf{b}_{2I+a}= \sum_E \int_{\Gamma\indices{_{[E]}}} \gamma \nabla\indices{_b}(N\indices{_{[E,I]}}  T\indices{_{[E,I]}^A_a}) g\indices{^{ab}}  dS .
\end{equation}
To simplify notation, we have defined the symbol
\begin{equation}
\nabla\indices{_b}(N\indices{_{[E,I]}} T\indices{_{[E,I]}^A_a}) = \partial\indices{_b}N\indices{_{[E,I]}} T\indices{_{[E,I]}^A_a} + N\indices{_{[E,I]}}\partial\indices{_b}T\indices{_{[E,I]}^A_a}-\Gamma^c_{ba} N\indices{_{[E,I]}} T\indices{_{[E,I]}^A_c},
\end{equation}
where $\Gamma\indices{^{a}_{bc}}=\partial\indices{_b} \partial\indices{_a}\bm{\psi}\indices{_{[E]}} \cdot\partial\indices{_c}\bm{\psi}\indices{_{[E]}}$ are the Christoffel symbols on $\Gamma$ in the basis $\mathcal{B}_{\bm{\psi}_{[E]}}$. 
The derivative $\partial\indices{_b}T\indices{_{[E,I]}^A_a}$ can be computed as
\begin{equation}
\partial\indices{_b}T\indices{_{[E,I]}^A_a}= \bm{i}\indices{_{[I]}_{A}}\cdot\partial\indices{_b} \partial\indices{_a}\bm{\psi}\indices{_{[E]}}.
\end{equation}
On a sphere, for $\gamma=z$ and $\mu=\eta=1$, the solution to Eq.~\eqref{marangoni} is $\bm{v}(\bm{x}) = \left(x \sqrt{1-\rho^2},y \sqrt{1-\rho^2}, -\rho^2\right)$, where $\rho=\sqrt{x^2+y^2}$. In Fig.~\ref{figmarfl} we plot the convergence rate to the analytical solution of the approximated vector with the LMP and the HD methods respectively, finding equivalent results to those in the previous section.

\tblue{
To further show the ability of the method to deal with vector-valued PDEs on different topologies, we now solve Eq.~\eqref{marangoni} on a double torus. We describe the double torus with an implicit equation given by $f_{tt}(\bm{x}) = f_{t_1}(\bm{x}) f_{t_2}(\bm{x}) -\epsilon = 0$, where $f_{t_1}=|\bm{x}|^2+R^2-r^2-2R\sqrt{x^2+y^2}$ is the implicit function of a torus centered at $(0,0,0)$ of major radius $R$ in the $xy$-plane and minor radius $r$ and $f_{t_2}=|\bm{x}|^2+R^2-r^2-2R\sqrt{(x-c_2)^2+z^2}$ is the implicit function of a torus of the same dimension but translated by $(c,0,0)$ and rotated $90$ degrees around the $y$ axis. We choose $R=0.45$, $r=0.1$ and $c=1$ and $\epsilon=10^{-4}$; to fit the torus, we minimize $(f_\text{tt}(\bm{x}))^2$ with a Newton-Raphson iteration. We impose a tension $\gamma=\cos(3\pi x) + \exp(-y^2)$ and compute $\bm{v}$, see Fig.~\ref{figmarfl}C. Since an analytical solution to this problem is not available, we analyze here the convergence in the functional $\mathcal{R}=\mu \int_\Gamma \left\{|\nabla^S\bm{v}|^2dS + \eta |\bm{v}|^2 + \gamma \nabla\cdot \bm{v}\right\}dS$. We compute the absolute value of the difference of $\mathcal{R}$ between consecutively refined meshes with one of a further level of refinement, see Fig.~\ref{figmarfl}D (in red). We observe a convergence rate of $r=2$ in this norm, which, as expected, has one order of convergence smaller than the $L_2$ norm because it involves derivatives of $\bm{v}$. We also show (in blue) the convergence in this norm for the problem in Fig.~\ref{figmarfl} B for comparison; note that since $h_0$ is larger for the sphere than for the torus, the two curves appear shifted.
}

\subsection{Nematic gel on a surface}
Finally, we describe the application of the LMP method to the discretization of a nematic tensor on a surface. A nematic tensor $\bm{Q}$ is a traceless symmetric tensor, often represented by 
\begin{equation}
Q\indices{_{ab}}= S\left(p\indices{_a} p\indices{_b} - \frac{1}{2}\delta\indices{_{ab}} \right),
\end{equation}
where $S=\sqrt{2||\bm{Q}||^2}$ is the so-called nematic order parameter, with $||\bm{Q}|| = \sqrt{Q\indices{_{ab}}g\indices{^{ac}}g\indices{^{bd}}Q\indices{_{cd}}}$ the norm of $\bm{Q}$, and $\bm{p}$ identifies the unit eigenvector of eigenvalue $S$ of $\bm{Q}$ \cite{De_Gennes1995-la}. Physically, $\bm{Q}$ is used to characterize an ensemble of rods, e.g.~filaments in the actin cortex \cite{Salbreux2009-nc} or cells in epithelial tissues \cite{Popovic2017-yh, Mueller2019-za}, with average alignment given by $\bm{p}$ and with ordering measured by $S$. For $S=0$ the rods are randomly distributed whereas for $S=1$ they are completely aligned in the direction of $\bm{p}$. \tblue{Note that for $S=0$, $\bm{p}$ is ill-defined. A simplified evolution law for $\bm{Q}$ is given by 
\begin{equation}
\label{evlaw}
\partial_t \bm{Q} = -\frac{\chi_1}{\mu}  \bm{Q}- \frac{\chi_2}{\mu} S^2  \bm{Q}+ \frac{L}{\mu} \Delta\bm{Q},
\end{equation}
where $\chi_1$ is the inverse susceptibility around the disordered state, $\chi_2$ its second-order correction, $L$ is the elastic Frank constant in the so-called one-constant approximation, $\mu$ measures the rotational viscosity, and $\Delta=\nabla\cdot\nabla$. A more elaborate version of this model was consistently derived in \cite{Nitschke2018-ru} as a thin-film limit of the Landau-de Gennes model \cite{De_Gennes1995-la}, and was later numerically exercised in \cite{Nestler2019-gk} using a cartesian representation of the tensor. For simplicity, here we neglect the coupling between $\bm{Q}$ and $\bm{k}$, the curvature tensor on the surface, which appears naturally in \cite{Nitschke2018-ru} when taking the limit of a thin film.
For positive $\chi_1$ and $\chi_2$, the disordered state ($S=0$) is at equilibrium; however, for $\chi_1<0$ and $\chi_2>0$ this models leads to a preferred nematic order given by $S_0=\sqrt{-\chi_1/(2\chi_2)}$. The evolution law in Eq.~\eqref{evlaw} can be shown to derive from a variational principle, which we state next.} 
\begin{figure}
	\begin{center}
		\includegraphics[width=6.5in]{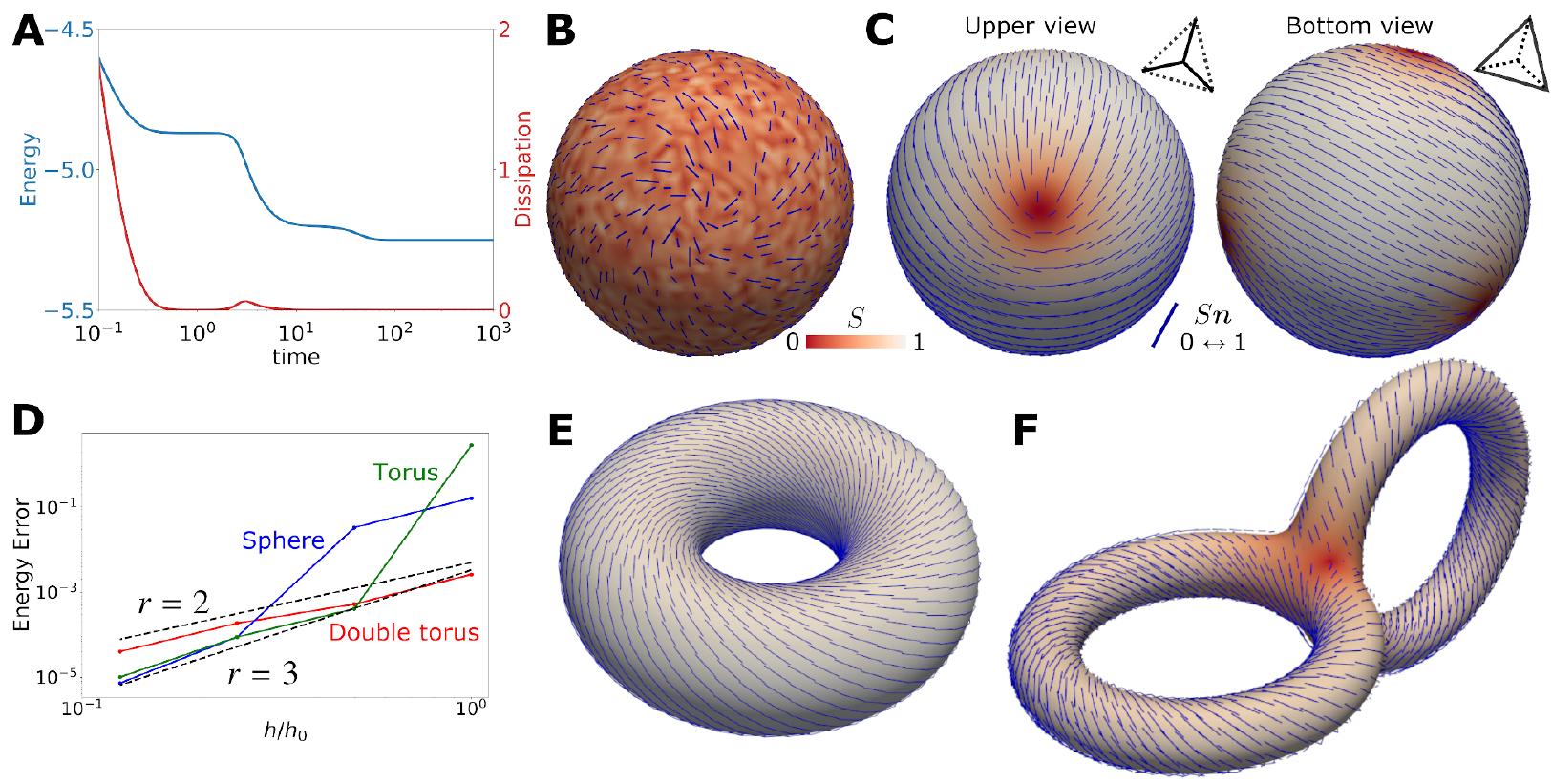}
		\caption{\label{fignem}\tblue{(A-C) Dynamics of a nematic system with $S_0=1$ on a sphere. Starting with a random configuration both for $S$ and $\bm{p}$ (B), we let the system evolve towards the configuration of minimal energy, formed by four +1/2 defects on the vertices of a tetrahedron (C). We observe that the free energy decreases along the dynamics (A), as expected from Onsager's formalism and the variational time-integrator. (D) Convergence in the energy norm of the final configuration for the sphere as well as for a torus (E), where there are no defects, and a double torus (F), where there are four -1/2 defects.}}
	\end{center}
\end{figure}
\tblue{First, we define the free energy of the system as
\begin{equation}
\label{energynematic}
\mathcal{F}[\bm{Q}] = \frac{1}{2}\int_{\Gamma}\left\{ \chi_1 S^2 + \chi_2 S^4 + L ||\nabla\bm{Q}||^2 \right\} dS,
\end{equation}
where $||\nabla\bm{Q}||=\sqrt{\nabla\bm{Q}\threedotsord\nabla\bm{Q}}$ and $\nabla\bm{Q}\threedotsord\nabla\bm{R} = \nabla_cQ_{ab} g^{ad}g^{be}g^{cf}\nabla_fR_{de}$.
The first two terms in Eq.~\eqref{energynematic} penalize deviations of $S$ from $S_0$ (as a function of $S^2$, the sum of these two terms looks like a parabola with minimum at $S_0^2$). The last term penalizes gradients in $\bm{Q}$.} We also define a dissipation potential, depending on the rate of change of $\bm{Q}$, as
\begin{equation}
\mathcal{D}[\partial_t\bm{Q}] = \frac{\mu}{2} \int_{\Gamma}||\partial_t\bm{Q}||^2 dS.
\end{equation}
Then, by defining the Rayleighian
\begin{equation}
\mathcal{R}[\bm{Q},\partial_t\bm{Q}]  = \partial_t\mathcal{F}[\bm{Q},\partial_t\bm{Q}] +\mathcal{D}[\partial_t\bm{Q}],
\end{equation}
one can show that the Euler-Lagrange equations associated to minimizing this functional,
\begin{equation}
\label{weaknem}
\partial_t\bm{Q} = \underset{\bm{R}}{\arg\min}~ \mathcal{R}[\bm{Q},\bm{R}],
\end{equation}
are precisely Eq.~\eqref{evlaw}. \tblue{Note that when taking the time derivative of Eq.~\eqref{energynematic}, $\partial_t\mathcal{F}[\bm{Q},\partial_t\bm{Q}]$, the result will in general depend on both $\bm{Q}$ and $\partial_t\bm{Q}$.}
To solve this problem numerically, we resort to the variational time-integrator developed in \cite{Torres-Sanchez2019-kt}, in which we minimize the discrete Rayleghian
\begin{equation}
\label{problemnem}
\mathcal{R}^n[\bm{Q}^n,\bm{Q}^{n+1}] = \frac{\mathcal{F}[\bm{Q}^{n+1}]-\mathcal{F}[\bm{Q}^{n}]}{\Delta t ^n} + \mathcal{D}\left[\frac{\bm{Q}^{n+1}-\bm{Q}^n}{\Delta t^n}\right],
\end{equation}
where $\Delta t^n$ is the (possibly non-uniform) time-step. As shown in \cite{Torres-Sanchez2019-kt}, this variational time-integrator is unconditionally stable since the free energy is a Lyapunov function of the discrete dynamics, as in the continuous case. In space, we discretize $\bm{Q}$ with a variation of the LMP method, which imposes by construction that $\bm{Q}$ is symmetric and traceless. We have two possible options for the discretization of $\bm{Q}$. On the one hand, if we discretize contravariant or covariant components of $\bm{Q}$, then symmetry can be simply imposed by imposing for each node $I$ that $Q\indices{^I_{21}} = Q\indices{^I_{12}}$, thus discarding $Q\indices{^I_{12}}$ from the degrees of freedom. Imposing that the trace of $\bm{Q}^I$ is zero using covariant components results in the  constraint $g^{ab}(\bm{x})Q_{ab}(\bm{x}) = 0$, which depends on the metric tensor. On the other hand, if we discretize mixed components of $\bm{Q}$, then the traceless constraint is simply $Q\indices{^1_1} + Q\indices{^2_2} = 0$, but imposing symmetry also involves $\bm{g}$. Here, we discretize covariant components of $\bm{Q}$, and construct traceless and symmetric nodal nematic tensors $\bm{Q}_{[I]}(\bm{x})$ as follows. 
We first represent $\bm{Q}_{[I]}(\bm{x})$ in the basis $\mathcal{B}_{\bm{\phi}_{[I]}}$ of the local Monge parametrization as
\begin{equation}
\left[\bm{Q}_{[I]}(\bm{x})\right]_{\mathcal{B}_{\bm{\phi}_{[I]}}} 
= \begin{pmatrix}
Q\indices{_{[I]}_{1}} & Q\indices{_{[I]}_{2}}\\
Q\indices{_{[I]}_{2}} & \displaystyle -\frac{g\indices{^{11}}(\bm{x})}{g\indices{^{22}}(\bm{x})}Q\indices{_{[I]}_{1}}-2\frac{g\indices{^{12}}(\bm{x})}{g\indices{^{22}}(\bm{x})}Q\indices{_{[I]}_{2}} \end{pmatrix}
=
Q\indices{_{[I]}_{1}} 
\begin{pmatrix}
1&0\\
0 & -\frac{g\indices{^{11}}(\bm{x})}{g\indices{^{22}}(\bm{x})}
\end{pmatrix} + Q\indices{_{[I]}_{2}} 
\begin{pmatrix}
0&1\\
1 & -2\frac{g\indices{^{12}}(\bm{x})}{g\indices{^{22}}(\bm{x})}
\end{pmatrix}
,
\end{equation}
where obviously the components of $\bm{g}$ are those in  $\mathcal{B}_{\bm{\phi}_{[I]}}$ and $Q\indices{_{[I]}_{C}}, ~C=1,2$, are the nodal coefficients. It is straightforward to see that this tensor is symmetric and traceless. We can express this equation as
\begin{equation}
\label{QI}
Q\indices{_{[I]}_{AB}}(\bm{x}) = L\indices{_{[I]}_{AB}^C}(\bm{x})Q\indices{_{[I]}_C},
\end{equation}
where $L\indices{_{[I]}_{AB}^C}$ is the third order symbol defined in matrix notation by
\begin{equation}
\left[L\indices{_{[I]}_{AB}^C}(\bm{x})\right] = \begin{pmatrix} 
\begin{pmatrix}
\displaystyle1&0\\
0 & 1
\end{pmatrix} 
\begin{pmatrix}
\displaystyle0&1\\
-\frac{g\indices{^{11}}(\bm{x})}{g\indices{^{22}}(\bm{x})} & -2\frac{g\indices{^{12}}(\bm{x})}{g\indices{^{22}}(\bm{x})}
\end{pmatrix}
\end{pmatrix}.
\end{equation}
Transforming the components of $\bm{Q}_{[I]}(\bm{x})$ from the basis $\mathcal{B}_{\bm{\phi}_{[I]}}$ to the natural finite element basis $\mathcal{B}_{\bm{\psi}_{[E]}}$ as previously described and then using the general LMP representation of tensors, we obtain
\begin{equation}
Q\indices{_{ab}}(\bm{x}) = \sum_{I\in\langle E\rangle}  Q\indices{_{[I]}_C} L\indices{_{[I]}_{AB}^C}(\bm{x}) \left(N\indices{_{[E,I]}} T\indices{_{[E,I]}^{A}_{a}} T\indices{_{[E,I]}^{B}_{b}}\right)\circ \bm{\psi}\indices*{_{[E]}^{-1}}(\bm{x}).
\end{equation}
To compute the covariant derivative of $\bm{Q}$ in the free energy, we apply the definition
\begin{equation}
\nabla\indices{_c} Q\indices{_{ab}} = \partial\indices{_c} Q\indices{_{ab}} - \Gamma\indices{^d_{ca}} Q_{db} - \Gamma\indices{^d_{ca}} Q_{ad},
\end{equation}
and
\begin{equation}
\begin{aligned}
\partial_c Q_{ab} =  \sum_{I\in\langle E\rangle} &\left\{\partial_cN\indices{_{[E,I]}} T\indices{_{[E,I]}^{A}_{a}} T\indices{_{[E,I]}^{B}_{b}} L\indices{_{[I]}_{AB}^C} Q\indices{_{[I]}_C} + N\indices{_{[E,I]}} \partial\indices{_c}T\indices{_{[E,I]}^{A}_{a}} T\indices{_{[E,I]}^{B}_{b}}  L\indices{_{[I]}_{AB}^C} Q\indices{_{[I]}_C} \right.\\
&\left.+\,N\indices{_{[E,I]}} T\indices{_{[E,I]}^{A}_{a}} \partial\indices{_c}T\indices{_{[E,I]}^{B}_{b}}  L\indices{_{[I]}_{AB}^C} Q\indices{_{[I]}_C} +N\indices{_{[E,I]}} T\indices{_{[E,I]}^{A}_{a}} T\indices{_{[E,I]}^{B}_{b}}\partial\indices{_c}  L\indices{_{[I]}_{AB}^C} Q\indices{_{[I]}_C}\right\},
\end{aligned}
\end{equation}
where
\begin{equation}
\partial_c  L\indices{_{[I]}_{11}^1} =  \frac{\partial_c g\indices{^{22}} g\indices{^{11}} - \partial_c g\indices{^{11}} g\indices{^{22}}}{\left(g\indices{^{22}}\right)^2},\quad \partial_c  L\indices{_{[I]}_{11}^2} = 2 \frac{\partial_c g\indices{^{22}} g\indices{^{12}} - \partial_c g\indices{^{12}} g\indices{^{22}}}{\left(g\indices{^{22}}\right)^2},\quad  \partial_c  L\indices{_{[I]}_{AB}^C} = 0 \text{ otherwise}.
\end{equation}
To calculate $\partial\indices{_c} g\indices{^{AB}}$ we note that
\begin{equation}
g\indices{^{AB}} = T\indices{_{[E,I]}^A_a} T\indices{_{[E,I]}^B_b} g^{ab} ,
\end{equation}
and thus
\begin{equation}
\partial\indices{_c} g\indices{^{AB}} = \partial\indices{_c}T\indices{_{[E,I]}^A_a} T\indices{_{[E,I]}^B_b} g^{ab} + T\indices{_{[E,I]}^A_a} \partial\indices{_c}T\indices{_{[E,I]}^B_b} g^{ab} + T\indices{_{[E,I]}^A_a} T\indices{_{[E,I]}^B_b} \partial\indices{_c}g^{ab}.
\end{equation}
Finally, we have that
\begin{equation}
\partial\indices{_c}g^{ab} = - g^{ad}g^{be} \partial\indices{_c}g_{de},
\end{equation}
and
\begin{equation}
\partial\indices{_c}g_{ab} = \partial\indices{_c}\left(\partial\indices{_a}\bm{\psi}\indices{_{[E]}}\cdot \partial\indices{_b}\bm{\psi}\indices{_{[E]}}\right) = \partial\indices{_c}\partial\indices{_a}\bm{\psi}\indices{_{[E]}} \cdot \partial\indices{_b}\bm{\psi}\indices{_{[E]}} + \partial\indices{_a}\bm{\psi}\indices{_{[E]}} \cdot \partial\indices{_c}\partial\indices{_b}\bm{\psi}\indices{_{[E]}}.
\end{equation}
To solve the non-linear problem resulting from this discretization and Eq.~\eqref{problemnem}, we use a Newton-Raphson method. 

\tblue{Taking $\chi_1=-1$, $\chi_2=2$, $L=1/10$ and $\mu=1$, we now examine the dynamics of the model simulated with the LMP method, see Fig.~\ref{fignem}. On a sphere, starting with a random configuration for $\bm{p}$ and $S$ (Fig.~\ref{fignem}B) we let the system relax its free energy (Fig.~\ref{fignem}A). The system tries to reach a state where $S=S_0=1$ everywhere to minimize the first part of the free energy, comprising the first two contributions in Eq.~\eqref{energynematic}. It also tries to reach a state where $\nabla\bm{Q}$ is small to minimize the last contribution. On a sphere, however, due to the hairy ball theorem, it is impossible to realize a homogeneous $\bm{Q}$ with $S=1$ that would lead to a vanishing free energy. The nematic field has to satisfy the Poincar\'e-Hopf theorem, which states that the sum of the charges of its zeros (or defects) has to be equal to the Euler characteristic of the surface, $2$ for a sphere.
In our simulations, the system relaxes towards a state in which four defects (of $+1/2$ charge and with a length-scale given by $\sqrt{\chi_2/L}$) are created forming a tetrahedron, as expected from classical theories \cite{Lubensky1992-zj} and recent experiments \cite{Keber2014-nv}, see Fig.~\ref{fignem}C. Looking at the same dynamics on a torus, however, we observe that the system relaxes towards a well-combed nematic phase with no defects, see Fig.~\ref{fignem}E, since the torus has an Euler characteristic of zero. On the double-torus, with an Euler characteristic of $-2$, the system develops four defects again, of $-1/2$ charge this time, that sit around the junction between the two tori, see Fig.~\ref{fignem}F. To show the convergence of the LMP method for this problem, we compute the energy of the final (stable) configuration of the dynamics for the different topologies. We calculate the absolute value of the difference between the energy for a given mesh and the energy for a further subdivided mesh, see Fig.~\ref{fignem}D. We observe  a second-order convergence on the sphere and the double torus and a third order convergence on the torus, which agree with the results shown in previous sections.}

\section{Discussion}

We have presented a novel method, the Local Monge Parametrizations (LMP) method, for the discretization of differentiable tensor fields on surfaces of arbitrary topology based only on tangential calculus. We have shown the applicability of the LMP method to approximate vector and second-order tensor fields on surfaces of different topology in combination with subdivision surface finite elements. We have also used the LMP method to solve two different vector- and tensor-valued PDEs, representing a surface-driven flow and a nematic system, on surfaces. 
\tblue{The main advantages of the LMP method with respect to previous methods are that it (1) is general with regards to the topology of the surface as opposed to methods based on the HD, (2) does not increase the order of vector- and tensor-valued PDEs, and (3) does not increase the number of degrees of freedom involved in the problem as opposed to Cartesian discretizations requiring tangency constraints such as in \cite{Nestler2019-gk}. The third point is specially relevant when dealing with higher order tensors, since the number of out-of-plane components of the tensor that are unessential and must be constrained increase dramatically with the order of the tensor: 1 for a vector, 5 for a second order tensor, 19 for a third order tensor, and so on. 
The LMP method leads, by construction, to a tangent vector/tensor field on the discrete surface. This differs from the Cartesian approach, where the tangency constraint is imposed weakly, usually through a penalty function \cite{Nestler2019-gk}. Thus, the results of the LMP do not depend on any penalty constant. On the other hand, a drawback is that if the tensor field needs to be differentiable, e.g.~to solve a second-order PDE, one needs to have a $C^1$ description of the surface. Here, we resorted to smooth Loop subdivision surfaces. A method based on a Cartesian approach does not have this requirement because the tensor field is not exactly tangent to the surface, and one can use a  $C^0$ discretization of the surface such as that obtained with classical FEM with Lagrange polynomials. The Cartesian method in \cite{Nestler2019-gk} requires an approximation to the curvature tensor on the $C^0$ surface. As opposed to usual FEM, the LMP method requires computing the matrices $T\indices{_{[E,I]}^A_a}$ (and outer products of them), which has a larger computational cost as compared to simply evaluating basis functions. Practically, however, this is not a limiting factor. On the one hand, for a time-evolution problem such as the last example, the matrices can be computed at the beginning of the simulation and used during the whole dynamics
On the other hand, in all problems considered here, we have found that, for relatively large meshes, the computational time required for solving the system of equations (either with an iterative solver such as GMRES with domain decomposition or with a parallel direct solver such as MUMPS) is always larger than the time required to assemble the matrix.} 
The LMP method can be used to discretize the models in \cite{Sahu2017-nz,Sauer2017-oc,Torres-Sanchez2019-kt,Salbreux2017-ln} involving  vector- and tensor-valued PDEs coupled to surface evolution laws. 
\tblue{On a time-evolving surface, $\Gamma_t$, an extension of the method should provide an algorithm to evolve the local parametrization around each node.
If the surface does not present large deformations, one could consider a fixed frame $\{\bm{i}\indices{_{[I]}_1},\bm{i}\indices{_{[I]}_2},\bm{n}\indices{_{[I]}}\}$. As long as the time-evolving neighborhood of $I$, $\Gamma_{t;[I]}$, has a well-defined projection onto $\Theta_{[I]}$, this method could be used; the transformation matrix $T\indices{_{t;[E,I]}^A_a}=\bm{i}\indices{_{[I]}_A}\cdot\partial_a\psi(t)_{[E]a}$ evolves following the evolution of $\bm{\psi}$.
 On the other hand, there are relevant examples, such as in a Lagrangian scheme or when the surface is parametrized through an offset \cite{Torres-Sanchez2019-kt}, where there is a notion of reference surface. Then, one has a mapping from the reference surface $\Gamma_0$ to the deformed surface, $\bm{\zeta}_t:\Gamma_0\rightarrow\Gamma_t$. One can then define a local time-evolving parametrization around node $I$ as $\bm{\phi}_{t;[I]}=\bm{\zeta}_t\circ\bm{\phi}_{[I]0}$ thus deforming the initial Monge parametrization $\bm{\phi}_{[I]0}$ with the time-evolution of the surface; in that case one has $T\indices{_{t;[E,I]}^A_a} = T\indices{_{0;[E,I]}^A_b}\partial\indices{_a}\zeta\indices{_t^b}$. Finally, one could devise other kinds of time-evolution for $\{\bm{i}\indices{_{[I]}_1},\bm{i}\indices{_{[I]}_2},\bm{n}\indices{_{[I]}}\}$ . For instance, one could dynamically evolve $\bm{n}_{[I]}$ to be the normal of the surface at the position of node $I$ and choose $\{\bm{i}\indices{_{[I]}_1},\bm{i}\indices{_{[I]}_2}\}$ perpendicular to $\bm{n}_{[I]}$ with an evolution law minimizing  $||\partial_t\bm{i}\indices{_{[I]}_i}||^2$.}
Our method could also find application in the discretization of smooth tensor fields in computer graphics  as an alternative to existing methods \cite{De_Goes2016-dd}. 

\section{Acknowledgments}

We acknowledge the support of the European Research Council (CoG-681434), the Spanish Ministry of Economy and Competitiveness/FEDER (DPI2015-71789-R to MA), and the Generalitat de Catalunya (SGR-1471, ICREA Academia award to MA). We also thank Sohan Kale and Guillermo Vilanova for useful discussions.

\appendix
\section{Change of basis}
\label{change-basis}
To compute the matrix for the change of basis, we note that it should satisfy
\begin{equation}
\label{app1}
T\indices{_{[E,I]}^A_a} \partial_A\bm{\phi}_{[I]} = \partial_a\bm{\psi}\indices*{_{[E]}}.
\end{equation}
We can use the definition of $\bm{\phi}_{[I]}$ Eq.~\eqref{eqlocmonpar} to compute
\begin{equation}
\partial_A \bm{\phi}\indices{_{[I]}} = \bm{i}\indices{_{[I]}_A} + \partial_A h\indices{_{[I]}} \bm{n}_{[I]}.
\end{equation}
It is easy to note that the dual basis of $\left\{\partial_A \bm{\phi}\indices{_{[I]}}\right\}$ is simply given by $\{\bm{i}\indices{_{[I]}_A}\}$, since
\begin{equation}
\bm{i}\indices{_{[I]}_B} \cdot \partial_A \bm{\phi}\indices{_{[I]}} = \delta_{AB}.
\end{equation}
Thus, multiplying Eq.~\eqref{app1} by $\bm{i}\indices{_{[I]}_B}$, we get
\begin{equation}
T\indices{_{[E,I]}^A_a} \bm{i}\indices{_{[I]}_B}\cdot \partial_A\bm{\phi}_{[I]} = T\indices{_{[E,I]}^A_a} \delta_{AB} = \bm{i}\indices{_{[I]}_B}\cdot \partial_a\bm{\psi}\indices*{_{[E]}}.
\end{equation}
which leads to Eq.~\eqref{transform1}.

\bibliographystyle{elsarticle-num}
\bibliography{ref}

\begin{thebibliography}{10}
\expandafter\ifx\csname url\endcsname\relax
  \def\url#1{\texttt{#1}}\fi
\expandafter\ifx\csname urlprefix\endcsname\relax\def\urlprefix{URL }\fi
\expandafter\ifx\csname href\endcsname\relax
  \def\href#1#2{#2} \def\path#1{#1}\fi

\bibitem{Marsden1994-lj}
J.~E. Marsden, T.~J.~R. Hughes, {Mathematical Foundations of Elasticity},
  Courier Corporation, New York, New York, USA, 1994.

\bibitem{Jackson2007-nf}
J.~D. Jackson, {Classical electrodynamics}, John Wiley \& Sons, 2007.

\bibitem{Holzapfel2000-go}
G.~A. Holzapfel, {Nonlinear Solid Mechanics: A Continuum Approach for
  Engineering}, Wiley, 2000.

\bibitem{De_Gennes1995-la}
P.~G. de~Gennes, J.~Prost, {The Physics of Liquid Crystals}, Clarendon Press,
  1995.

\bibitem{Doi2013-dq}
M.~Doi, {Soft Matter Physics}, OUP Oxford, 2013.

\bibitem{Rahimi2012-sa}
M.~Rahimi, M.~Arroyo, {Shape dynamics, lipid hydrodynamics, and the complex
  viscoelasticity of bilayer membranes}, Physical Review E 86~(1) (2012)
  011932.

\bibitem{Arroyo2009-xz}
M.~Arroyo, A.~Desimone, {Relaxation dynamics of fluid membranes}, Phys. Rev. E
  Stat. Nonlin. Soft Matter Phys. 79~(3 Pt 1) (2009) 031915.

\bibitem{Sahu2017-nz}
A.~Sahu, R.~A. Sauer, K.~K. Mandadapu, {Irreversible thermodynamics of curved
  lipid membranes}, Phys Rev E 96~(4-1) (2017) 042409.

\bibitem{Sauer2017-oc}
R.~A. Sauer, T.~X. Duong, K.~Mandadapu, D.~Steigmann, {A stabilized finite
  element formulation for liquid shells and its application to lipid bilayers},
  J. Comput. Phys. 330 (2017) 1--19.

\bibitem{Torres-Sanchez2019-kt}
A.~Torres-S{\'a}nchez, D.~Mill{\'a}n, M.~Arroyo, {Modelling fluid deformable
  surfaces with an emphasis on biological interfaces}, J. Fluid Mech. 872
  (2019) 218--271.

\bibitem{Turlier2014-wh}
H.~Turlier, B.~Audoly, J.~Prost, J.-F. Joanny, {Furrow constriction in animal
  cell cytokinesis}, Biophys. J. 106~(1) (2014) 114--123.

\bibitem{Salbreux2017-ln}
G.~Salbreux, F.~J{\"u}licher, {Mechanics of active surfaces}, Phys Rev E
  96~(3-1) (2017) 032404.

\bibitem{Mietke2019-fe}
A.~Mietke, F.~J{\"u}licher, I.~F. Sbalzarini, {Self-organized shape dynamics of
  active surfaces}, Proc. Natl. Acad. Sci. U. S. A. 116~(1) (2019) 29--34.

\bibitem{Ishihara2017-pn}
S.~Ishihara, P.~Marcq, K.~Sugimura, {From cells to tissue: A continuum model of
  epithelial mechanics}, Phys Rev E 96~(2-1) (2017) 022418.

\bibitem{Latorre2018-cc}
E.~Latorre, S.~Kale, L.~Casares, M.~G{\'o}mez-Gonz{\'a}lez, M.~Uroz, L.~Valon,
  R.~V. Nair, E.~Garreta, N.~Montserrat, A.~Del~Campo, B.~Ladoux, M.~Arroyo,
  X.~Trepat, {Active superelasticity in three-dimensional epithelia of
  controlled shape}, Nature 563~(7730) (2018) 203--208.

\bibitem{Nitschke2012-nh}
I.~Nitschke, A.~Voigt, J.~Wensch, {A finite element approach to incompressible
  two-phase flow on manifolds}, J. Fluid Mech. 708 (2012) 418--438.

\bibitem{Reuther2015-qv}
S.~Reuther, A.~Voigt, {The Interplay of Curvature and Vortices in Flow on
  Curved Surfaces}, Multiscale Model. Simul. 13~(2) (2015) 632--643.

\bibitem{Nitschke2017-pv}
I.~Nitschke, S.~Reuther, A.~Voigt, {Discrete Exterior Calculus (DEC) for the
  Surface Navier-Stokes Equation}, in: D.~Bothe, A.~Reusken (Eds.), {Transport
  Processes at Fluidic Interfaces}, Springer International Publishing, Cham,
  2017, pp. 177--197.

\bibitem{Reuther2018-ky}
S.~Reuther, A.~Voigt, {Solving the incompressible surface Navier-Stokes
  equation by surface finite elements}, Phys. Fluids 30~(1) (2018) 012107.

\bibitem{Hansbo2019-dy}
P.~Hansbo, M.~G. Larson, K.~Larsson, {Analysis of finite element methods for
  vector Laplacians on surfaces}, IMA J. Numer. Anal. (Apr. 2019).

\bibitem{Nitschke2018-ru}
I.~Nitschke, M.~Nestler, S.~Praetorius, H.~L{\"o}wen, A.~Voigt, {Nematic liquid
  crystals on curved surfaces: a thin film limit}, Proc. Royal Soc. A
  474~(2214) (2018) 20170686.

\bibitem{Praetorius2018-nx}
S.~Praetorius, A.~Voigt, R.~Wittkowski, H.~L{\"o}wen, {Active crystals on a
  sphere}, Phys Rev E 97~(5-1) (2018) 052615.

\bibitem{Nestler2018-el}
M.~Nestler, I.~Nitschke, S.~Praetorius, A.~Voigt, {Orientational Order on
  Surfaces: The Coupling of Topology, Geometry, and Dynamics}, J. Nonlinear
  Sci. 28~(1) (2018) 147--191.

\bibitem{Nitschke2019-hl}
I.~Nitschke, S.~Reuther, A.~Voigt, {Hydrodynamic interactions in polar liquid
  crystals on evolving surfaces}, Phys. Rev. Fluids 4~(4) (2019) 044002.

\bibitem{Sauer2019-dv}
R.~A. Sauer, R.~Ghaffari, A.~Gupta, {The multiplicative deformation split for
  shells with application to growth, chemical swelling, thermoelasticity,
  viscoelasticity and elastoplasticity}, Int. J. Solids Struct. 174-175 (2019)
  53--68.

\bibitem{Fisher2007-cn}
M.~Fisher, P.~Schr{\"o}der, M.~Desbrun, H.~Hoppe, {Design of Tangent Vector
  Fields}, in: {ACM SIGGRAPH 2007 Papers}, SIGGRAPH '07, ACM, New York, NY,
  USA, 2007.

\bibitem{Zhang2006-du}
E.~Zhang, K.~Mischaikow, G.~Turk, G.~Turk, {Vector Field Design on Surfaces},
  ACM Trans. Graph. 25~(4) (2006) 1294--1326.

\bibitem{Chen2007-uo}
G.~Chen, K.~Mischaikow, R.~S. Laramee, P.~Pilarczyk, E.~Zhang, {Vector field
  editing and periodic orbit extraction using Morse decomposition}, IEEE Trans.
  Vis. Comput. Graph. 13~(4) (2007) 769--785.

\bibitem{Palacios2007-cf}
J.~Palacios, E.~Zhang, {Rotational Symmetry Field Design on Surfaces}, ACM
  Trans. Graph. 26~(3) (Jul. 2007).

\bibitem{Polthier2003-mk}
K.~Polthier, E.~Preu{\ss}, {Identifying Vector Field Singularities Using a
  Discrete Hodge Decomposition}, in: {Visualization and Mathematics III},
  Springer Berlin Heidelberg, 2003, pp. 113--134.

\bibitem{Tong2003-sr}
Y.~Tong, S.~Lombeyda, A.~N. Hirani, M.~Desbrun, {Discrete Multiscale Vector
  Field Decomposition}, ACM Trans. Graph. 22~(3) (2003) 445--452.

\bibitem{Hirani2003-ho}
A.~N. Hirani, {Discrete exterior calculus}, Ph.D. thesis, California Institute
  of Technology (2003).

\bibitem{Wardetzky2006-gb}
M.~Wardetzky, {Discrete Differential Operators on Polyhedral Surfaces:
  Convergence and Approximation}, Ph.D. thesis, Freie Universit{\"a}t Berlin
  (2006).

\bibitem{Arnold2006-nk}
D.~N. Arnold, R.~S. Falk, R.~Winther, {Finite element exterior calculus,
  homological techniques, and applications}, Acta Numer. 15 (2006) 1--155.

\bibitem{Desbrun2008-ot}
M.~Desbrun, E.~Kanso, Y.~Tong, {Discrete Differential Forms for Computational
  Modeling}, in: A.~I. Bobenko, J.~M. Sullivan, P.~Schr{\"o}der, G.~M. Ziegler
  (Eds.), {Discrete Differential Geometry}, Birkh{\"a}user Basel, Basel, 2008,
  pp. 287--324.

\bibitem{Azencot2013-nf}
O.~Azencot, M.~Ben-Chen, F.~Chazal, M.~Ovsjanikov, {An Operator Approach to
  Tangent Vector Field Processing}, in: {Proceedings of the Eleventh
  Eurographics/ACMSIGGRAPH Symposium on Geometry Processing}, SGP '13,
  Eurographics Association, Aire-la-Ville, Switzerland, Switzerland, 2013, pp.
  73--82.

\bibitem{Azencot2015-ho}
O.~Azencot, M.~Ovsjanikov, F.~Chazal, M.~Ben-Chen, {Discrete Derivatives of
  Vector Fields on Surfaces -- An Operator Approach}, ACM Trans. Graph. 34~(3)
  (2015) 29.

\bibitem{De_Goes2014-nj}
F.~de~Goes, B.~Liu, M.~Budninskiy, Y.~Tong, M.~Desbrun, {Discrete-Tensor Fields
  on Triangulations}, Eurographics Symposium on Geometry Processing 33~(5)
  (2014) 13--24.

\bibitem{Turk2001-ed}
G.~Turk, {Texture Synthesis on Surfaces}, in: {Proceedings of the 28th Annual
  Conference on Computer Graphics and Interactive Techniques}, SIGGRAPH '01,
  ACM, New York, NY, USA, 2001, pp. 347--354.

\bibitem{Wei2001-va}
L.-Y. Wei, M.~Levoy, {Texture Synthesis over Arbitrary Manifold Surfaces}, in:
  {Proceedings of the 28th Annual Conference on Computer Graphics and
  Interactive Techniques}, SIGGRAPH '01, ACM, New York, NY, USA, 2001, pp.
  355--360.

\bibitem{Hertzmann2000-pz}
A.~Hertzmann, D.~Zorin, {Illustrating Smooth Surfaces}, in: {Proceedings of the
  27th Annual Conference on Computer Graphics and Interactive Techniques},
  SIGGRAPH '00, ACM Press/Addison-Wesley Publishing Co., New York, NY, USA,
  2000, pp. 517--526.

\bibitem{Panozzo2014-bf}
D.~Panozzo, E.~Puppo, M.~Tarini, O.~Sorkine-Hornung, {Frame Fields: Anisotropic
  and Non-orthogonal Cross Fields}, ACM Trans. Graph. 33~(4) (2014)
  134:1--134:11.

\bibitem{Vaxman2016-tp}
A.~Vaxman, M.~Campen, O.~Diamanti, D.~Panozzo, D.~Bommes, K.~Hildebrandt,
  M.~Ben-Chen, {Directional Field Synthesis, Design, and Processing}, Comput.
  Graph. Forum 35~(2) (2016) 545--572.

\bibitem{Bhatia2013-ye}
H.~Bhatia, G.~Norgard, V.~Pascucci, P.-T. Bremer, {The Helmholtz-Hodge
  decomposition--a survey}, IEEE Trans. Vis. Comput. Graph. 19~(8) (2013)
  1386--1404.

\bibitem{Fries2018-lv}
T.-P. Fries, {Higher-order surface FEM for incompressible Navier-Stokes flows
  on manifolds}, Int. J. Numer. Methods Fluids 88~(2) (2018) 55--78.

\bibitem{Nestler2019-gk}
M.~Nestler, I.~Nitschke, A.~Voigt, {A finite element approach for vector- and
  tensor-valued surface PDEs}, J. Comput. Phys. 389 (2019) 48--61.

\bibitem{Liu2016-eq}
B.~Liu, Y.~Tong, F.~D. Goes, M.~Desbrun, {Discrete Connection and Covariant
  Derivative for Vector Field Analysis and Design}, ACM Trans. Graph. 35~(3)
  (2016) 23:1--23:17.

\bibitem{De_Goes2016-dd}
F.~de~Goes, M.~Desbrun, Y.~Tong, {Vector Field Processing on Triangle Meshes},
  in: {ACM SIGGRAPH 2016 Courses}, SIGGRAPH '16, ACM, New York, NY, USA, 2016,
  pp. 27:1--27:49.

\bibitem{Stam1999-fk}
J.~Stam, {Evaluation of Loop subdivision surfaces}, in: {SIGGRAPH'99 Course
  notes}, Los Angeles, CA, 1999.

\bibitem{Cirak2000-tq}
F.~Cirak, M.~Ortiz, {Schroder (2000) Subdivision surfaces: a new paradigm for
  thin shell finite-element analysis}, Int. J. Numer. Methods Eng. 47~(12)
  (2000) 2039--2072.

\bibitem{Milnor1978-gf}
J.~Milnor, {Analytic Proofs of the ``Hairy Ball Theorem'' and the Brouwer Fixed
  Point Theorem}, Am. Math. Mon. 85~(7) (1978) 521--524.

\bibitem{Secomb1982-cf}
T.~W. Secomb, R.~Skalak, {Surface flow of viscoelastic membranes in viscous
  fluids}, Quart. J. Mech. Appl. Math. 35~(2) (1982) 233--247.

\bibitem{Morris2015-up}
R.~G. Morris, M.~S. Turner, {Mobility Measurements Probe Conformational Changes
  in Membrane Proteins due to Tension}, Phys. Rev. Lett. 115~(19) (2015)
  198101.

\bibitem{Sigurdsson2016-jt}
J.~K. Sigurdsson, P.~J. Atzberger, {Hydrodynamic coupling of particle
  inclusions embedded in curved lipid bilayer membranes}, Soft Matter 12~(32)
  (2016) 6685--6707.

\bibitem{Gross2018-nt}
B.~J. Gross, P.~J. Atzberger, {Hydrodynamic flows on curved surfaces: Spectral
  numerical methods for radial manifold shapes}, J. Comput. Phys. 371 (2018)
  663--689.

\bibitem{Mickelin2018-ug}
O.~Mickelin, J.~S{\l}omka, K.~J. Burns, D.~Lecoanet, G.~M. Vasil, L.~M. Faria,
  J.~Dunkel, {Anomalous Chained Turbulence in Actively Driven Flows on
  Spheres}, Phys. Rev. Lett. 120~(16) (2018) 164503.

\bibitem{Do_Carmo1992-bx}
M.~P. do~Carmo, {Riemannian geometry}, Vol. 115, Birkh{\"a}user Boston, Boston,
  1992.

\bibitem{Biermann2000-bz}
H.~Biermann, A.~Levin, D.~Zorin, {Piecewise smooth subdivision surfaces with
  normal control}, in: {Proceedings of the 27th annual conference on Computer
  graphics and interactive techniques}, ACM Press/Addison-Wesley Publishing
  Co., New York, New York, USA, 2000, pp. 113--120.

\bibitem{Arden2001-id}
G.~Arden, {Approximation properties of subdivision surfaces}, Ph.D. thesis,
  University of Washington (2001).

\bibitem{Salbreux2009-nc}
G.~Salbreux, J.~Prost, J.~F. Joanny, {Hydrodynamics of cellular cortical flows
  and the formation of contractile rings}, Phys. Rev. Lett. 103~(5) (2009)
  058102.

\bibitem{Popovic2017-yh}
M.~Popovi{\'c}, A.~Nandi, M.~Merkel, R.~Etournay, S.~Eaton, F.~J{\"u}licher,
  G.~Salbreux, {Active dynamics of tissue shear flow}, New J. Phys. 19~(3)
  (2017) 033006.

\bibitem{Mueller2019-za}
R.~Mueller, J.~M. Yeomans, A.~Doostmohammadi, {Emergence of Active Nematic
  Behavior in Monolayers of Isotropic Cells}, Phys. Rev. Lett. 122~(4) (2019)
  048004.

\bibitem{Lubensky1992-zj}
T.~C. Lubensky, J.~Prost, {Orientational order and vesicle shape}, J. Phys. II
  2~(3) (1992) 371--382.

\bibitem{Keber2014-nv}
F.~C. Keber, E.~Loiseau, T.~Sanchez, S.~J. DeCamp, L.~Giomi, M.~J. Bowick,
  M.~C. Marchetti, Z.~Dogic, A.~R. Bausch, {Topology and dynamics of active
  nematic vesicles}, Science 345~(6201) (2014) 1135--1139.

\end{thebibliography}

\end{document}